\documentclass[12pt]{article}
\usepackage[margin=1in]{geometry}
\usepackage{amsfonts,amssymb,epsfig,amsmath}
\usepackage{slashed}
\usepackage{color}
\usepackage{hyperref}

\renewcommand{\text}[1]{#1}

\newcommand{\be}{\begin{equation}}
\newcommand{\ee}{\end{equation}}
\newcommand{\ben}{\begin{displaymath}}
\newcommand{\een}{\end{displaymath}}
\newcommand{\bea}{\begin{eqnarray}}
\newcommand{\eea}{\end{eqnarray}}
\newcommand{\bean}{\begin{eqnarray*}}
\newcommand{\eean}{\end{eqnarray*}}
\newcommand{\nn}{\nonumber \\}
\newcommand{\ba}{\begin{array}}
\newcommand{\ea}{\end{array}}
\newcommand{\bi}{\begin{itemize}}
\newcommand{\ei}{\end{itemize}}

\renewcommand{\theequation}{\arabic{section}.\arabic{equation}}
\def\theequation{\thesection.\arabic{equation}}

\def\b{\beta}
\def\g{\gamma}

\def\g{\gamma}
\def\e{\epsilon}
\def\s{\sigma}
\def\e{\epsilon}

\renewcommand{\Box}{\square}

\def\mc{\mathcal}

\DeclareMathOperator{\vol}{vol}

\newcommand{\dd}{\mathrm{d}}
\newcommand{\DD}{\mathrm{D}}

\newcommand{\ft}[2]{{\textstyle\frac{#1}{#2}}}

\begin{document}

\makeatletter
\renewcommand{\theequation}{\thesection.\arabic{equation}}
\@addtoreset{equation}{section}
\makeatother

\begin{titlepage}

\vfill
\begin{flushright}
FPAUO-13/09\\
\end{flushright}

\vfill

\begin{center}
   \baselineskip=16pt
   {\Large \bf 3D supergravity from wrapped D3-branes}
   \vskip 2cm
     Parinya Karndumri$^{1,2}$ \& Eoin \'O Colg\'ain$^3$
       \vskip .6cm
             \begin{small}
               \textit{$^1$Department of Physics, Faculty of Science, Chulalongkorn University, \\ Bangkok 10330, THAILAND}
                 \vspace{2mm}

                 \textit{$^2$Thailand Center of Excellence in Physics, CHE, Ministry of Education,\\ Bangkok 10400, THAILAND}
                 \vspace{2mm}

      		 \textit{$^3$Departamento de F\'isica,
		 Universidad de Oviedo, \\
33007 Oviedo, ESPA\~NA}
             \end{small}\\*[.6cm]
\end{center}

\vfill \begin{center} \textbf{Abstract}\end{center} \begin{quote}
$AdS_3$ solutions dual to $\mathcal{N} = (0,2)$ SCFTs arise when D3-branes wrap K\"{a}hler two-cycles in manifolds with $SU(4)$ holonomy. Here we review known $AdS_3$ solutions and identify the corresponding three-dimensional gauged supergravities, solutions of which uplift to type IIB supergravity. In particular, we discuss gauged supergravities dual to twisted compactifications on Riemann surfaces of both $\mathcal{N}=4$ SYM and $\mathcal{N} =1$ SCFTs with Sasaki-Einstein duals. We check in each case that $c$-extremization gives the exact central charge and R symmetry. For completeness, we also study $AdS_3$ solutions from intersecting D3-branes, generalise recent KK reductions of Detournay \& Guica and identify the underlying gauged supergravities. Finally, we discuss examples of null-warped $AdS_3$ solutions to three-dimensional gauged supergravity, all of which embed in string theory.
\end{quote} \vfill

\end{titlepage}

\tableofcontents
\setcounter{table}{0}

\section{Introduction}
Gauged supergravity is a very useful tool in many areas of string theory such as flux compactifications and the AdS/CFT correspondence (see \cite{Samtleben:lecture} for a review). Due to these applications, gauged supergravities in various dimensions as well as their Kaluza-Klein (KK) dimensional reductions have been extensively explored. It is well known that lower-dimensional gauged supergravities can be obtained from dimensional reductions of higher-dimensional theories. Up to now, many examples have appeared and amongst them, \cite{dewit}, \cite{s41} and \cite{s51} are recognizable primary examples. In this paper, we are interested in gauged supergravities in three dimensions in order to incorporate both the principle of $c$-extremization and null-warped $AdS_3$ solutions.

The complete classification of Chern-Simons gauged supergravities in three dimensions has been given in \cite{deWit:2003ja}. Most theories constructed in this formulation have no known higher-dimensional origin. The three-dimensional gauged supergravities obtainable from dimensional reductions form a small part, with non-semisimple gauge groups, in this classification \cite{Nicolai:2003bp}. Unlike in higher-dimensional analogues, only a few examples of three-dimensional gauged supergravities, which play an important role in AdS$_3$/CFT$_2$ correspondence, have been obtained by dimensional reductions \cite{Colgain:2010rg, Karndumri:2010, Pope:SU2reduction}. In this paper, we will extend this list with more examples of gauged supergravities in three dimensions arising from wrapped D3-branes in type IIB supergravity.

Recently, $c$-extremization for $\mc{N}=(0,2)$ two-dimensional SCFT's has been proposed and various examples of gravity duals in five- and seven-dimensional gauged supergravities exhibited \cite{Benini:2012cz, Benini:2013cda}. Recall that $c$-extremization is a procedure that allows one to single out the correct $U(1)_R$ symmetry of the CFT from the mixing with other $U(1)$ symmetries. Soon after,  $c$-extremization was formulated purely in the context of the AdS$_3$/CFT$_2$ correspondence by explicitly showing that, in the presence of a gauged $SO(2)_R\sim U(1)_R $ R symmetry, the so-called $T$ tensor of the three-dimensional gauged supergravity can be extremized leading to the exact central charge and R symmetry \cite{Karndumri:2013iqa}. This realization is similar to how $a$-maximization of four-dimensional SCFT's \cite{Intriligator:2003jj} can be encoded in five-dimensional gauged supergravity  \cite{Tachikawa:2005} in the context of the AdS$_5$/CFT$_4$ correspondence\footnote{A concrete realization is presented in \cite{Szepietowski:2012tb}. }. Interestingly, in three dimensions, not only is the central charge reproduced, but the moment maps comprising the $T$ tensor give information about the exact R symmetry. In this work we will provide more details of the results quoted in \cite{Karndumri:2013iqa} and also exhibit another (related) example by considering twists of generic SCFT's with Sasaki-Einstein duals.

In three dimensions, where a vector is dual to a scalar, the matter coupled supergravity theory can be formulated purely in terms of scalar fields resulting in a non-linear sigma model coupled to supergravity. $\mc{N}=2$ supersymmetry in three dimensions requires the scalar target manifold to be K\"ahler. Gaugings of the theory are implemented by the embedding tensor specifying the way in which the gauge group is embedded in the global symmetry group. In general, the moment map of the embedding tensor, given by scalar matrices $\mc{V}$,  determines the $T$ tensor which plays an important role in computing the scalar potential and supersymmetry transformations. As a general result, $\mc{N}=2$ supersymmetry allows any proper subgroup of the symmetry to be gauged. Furthermore, there is a possibility of other deformations through a holomorphic superpotential ${W}$. The scalar potential generally gets contributions from both the $T$ tensor and the superpotential. However, any gauging of the R symmetry requires vanishing ${W}$.



The particular higher-dimensional theories we choose to reduce can all be motivated from the perspective of ten dimensions. From either an analysis of the Killing spinor equations \cite{Kim:2005ez}, or by following wrapped D-brane intuition \cite{Gauntlett:2007ph}, it is known that supersymmetric $AdS_3$ solutions supported by the five-form RR flux of type IIB supergravity, or in other words, those corresponding to wrapped D3-branes, have seven-dimensional internal manifolds $Y_7$ and bear some resemblance to Sasaki-Einstein metrics. More precisely, $Y_7$ can be expressed locally in terms of a natural $U(1)$ fibration (the R symmetry) over a six-dimensional K\"{a}hler base that is subject to a single differential condition
\be
\label{diffM6}
\Box R = \tfrac{1}{2} R^2 - R_{ij} R^{ij},
\ee
where $R$ and $R_{ij}$ are, respectively, the Ricci scalar and Ricci tensor of the metric of the K\"{a}hler manifold.
Through the supersymmetry conditions \cite{Kim:2005ez, Gauntlett:2007ph}, the Ricci scalar $R$ is related to an overall warp factor for the ten-dimensional space-time.

Of course the above equation can be simplified considerably by assuming that the K\"{a}hler manifold is also Einstein, but in general, solutions with non-trivial warp factors can be difficult to find. A search for IIB solutions tailored to this context can be found in \cite{Gauntlett:2006ns}, where a solution originally found in \cite{Gauntlett:2006qw} was recovered. The challenges here are reminiscent of generalisations of direct-product $AdS_4$ and $AdS_5$ solutions to warped products. To support this observation, we recall that, for an Ansatz covering the most general supersymmetric warped $AdS_5$ solutions of type IIB supergravity \cite{Gauntlett:2005ww}, the only warped geometry\footnote{A class of solutions can be generated via TsT transformations \cite{Lunin:2005jy} starting from $AdS_5 \times S^5$, but as the transformation only acts on the internal $S^5$, the final solution is not warped.} noted by the authors beyond the special case of Sasaki-Einstein was the Pilch-Warner solution \cite{Pilch:2000ej}. On a more recent note, warped $AdS_4$ solutions of eleven-dimensional supergravity generalising Sasaki-Einstein have been found \cite{Gabella:2012rc, Halmagyi:2012ic}. In the face of these difficulties, it is a pleasant surprise to witness the ease at which supersymmetric solutions with warp factors can be constructed in five-dimensional supergravity through twisted compactifications on a constant curvature Riemann surface $\Sigma_{\frak{g}}$ of genus $\frak{g}$ and how the principle of $c$-extremization accounts for the central charge and exact R symmetry of the dual $\mathcal{N} = (0,2)$ SCFT \cite{Benini:2012cz, Benini:2013cda}.

$c$-extremization aside, we can further motivate the study of three-dimensional gauged supergravities through the continued interest in ``null warped" $AdS_3$ space-times. Over the last few years, we have witnessed a hive of activity surrounding warped $AdS_3$ space-times and their field theory duals \cite{Anninos:2008fx}, primarily in Topologically Massive Gravity (TMG) \cite{Deser:1982vy, Deser:1981wh}. Indeed, the mere existence of these solutions and the fact that they are deformations of $AdS_3$ with $SL(2,\mathbb{R}) \times U(1)$ isometry,  raises very natural questions about the putative dual CFT. Since relatively little is known about these theories, the common approach is to extract information holographically from warped $AdS_3$ solutions. To date, in three dimensions, warped $AdS_3$ solutions have cropped up in a host of diverse settings, including of course, solutions \cite{Anninos:2008fx,Moussa:2003fc, Bouchareb:2007yx} to TMG, solutions \cite{Clement:2009gq} to New Massive Gravity \cite{Bergshoeff:2009hq}, Higher-Spin Gravity \cite{Gary:2012ms}, topologically gauged CFTs \cite{Gran:2012mg} and three-dimensional gravity with a Chern-Simons (CS) Maxwell term \cite{Detournay:2012dz}, where the latter is embeddable in string theory.  As we shall see, within the last class of three-dimensional theories, one also finds gauged supergravities.

Indeed, ``null warped" $AdS_3$ are central to efforts to generalise AdS/CFT to a non-relativistic setting, where holography may be applicable to condensed matter theory via a class of Schr\"{o}dinger space-times. Taking the catalyst from \cite{Son:2008ye,Balasubramanian:2008dm}, through fledgling embeddings in string theory \cite{Herzog:2008wg, Maldacena:2008wh, Adams:2008wt, Hartnoll:2008rs}, various attempts have been made to provide a working description of non-relativistic holography.  On one hand, one may wish to start with a recognisable theory with Schr\"{o}dinger symmetry, such as a non-relativistic limit \cite{Nakayama:2009cz, Lee:2009mm} of ABJM \cite{Aharony:2008ug}, but holographic studies  \cite{Colgain:2009wm, Ooguri:2009cv, Jeong:2009aa} fail to capture the required high degree of supersymmetry. On the other hand, if one starts from gravity solutions with Schr\"{o}dinger symmetry, one may be more pragmatic and obtain an effective description of the dual non-relativistic CFT, valid at  large $N$ and strong coupling \cite{Guica:2010sw}\footnote{Separately it has been argued \cite{Janiszewski:2012nf, Janiszewski:2012nb} that generic non-relativistic quantum field theories have a holographic description in terms of Ho\v{r}ava gravity \cite{Horava:2009uw}.}.  Similar points of view were also advocated in \cite{Bobev:2011qx, Kraus:2011pf, ElShowk:2011cm}. Whether the dual theory is a genuine CFT as proposed in \cite{Anninos:2008fx}, or some warped CFT, is an open question drawing considerable attention\footnote{See \cite{Detournay:2012pc} for a recent discussion.}.

The structure of the rest of this paper is as follows. In section \ref{sec:wrappedD3}, we present an overview of our knowledge of supersymmetric $AdS_3$ geometries arising from wrapped D3-branes. In section \ref{sec:D3N02}, we focus on geometries with a $U(1)$ R symmetry dual to $\mathcal{N} = (0,2)$ SCFT's and present known examples preserving at least four supersymmetries, all of which will correspond to the vacua of the gauged supergravities we discuss later. In section \ref{sec:generic}, we provide more details of the KK reduction reported in \cite{Karndumri:2013iqa}. In section \ref{sec:N1twist}, we present the three-dimensional gauged supergravity corresponding to a twisted compactification of an $\mathcal{N}=1$ SCFT with a generic Sasaki-Einstein dual. In section \ref{sec:D3}, we generalise the KK reductions discussed in \cite{Detournay:2012dz} and identify the corresponding gauged supergravities. In section \ref{sec:Sch} we present some simple constructions of null-warped $AdS_3$, or alternatively Schr\"{o}dinger geometries with dynamical exponent $z=2$, before discussing some open avenues for future study in section \ref{sec:outlook}.

\section{$AdS_3$ from wrapped D3-branes}
\label{sec:wrappedD3}

\subsection{Review of wrapped D3-branes}
In this section we review supersymmetric $AdS_3$ geometries arising from D3-branes wrapping calibrated two-cycles in manifolds with $SU(2), SU(3)$ and $SU(4)$ holonomy. To this end, we follow the general ten-dimensional classification presented in \cite{Gauntlett:2007ph} and later indicate where particular explicit solutions fit into the bigger picture. The approach of \cite{Gauntlett:2007ph} builds on earlier work concerning wrapped M5-branes \cite{Gauntlett:2006ux,Figueras:2007cn}  and M2-branes \cite{MacConamhna:2006nb}.

We recall that the general ``wrapped-brane" strategy \cite{Gauntlett:2006ux} involves first assuming that $AdS_3$ geometries start off as warped products of the form
\be
\dd s^2_{10} = L^{-1} \dd s^2(\mathbb{R}^{1,1}) + \dd s^2(\mathcal{M}_8),
\ee
where both the warp factor $L$ and the metric on $\mathcal{M}_8$ are independent of the Minkowski factor. Here the Minkowski space-time should be regarded as the unwrapped part of the D3-brane, and as expected, the D3-branes source a self-dual RR five-form flux $F_5 = \Theta + *_{10} \Theta$ invariant under the symmetries of the Minkowski factor.

For the particular geometries of interest to us, the metric and the flux for the geometry may be expressed as \cite{Gauntlett:2007ph}
\bea
\dd s^2_{10} &=& L^{-1} \dd s^2(\mathbb{R}^{1,1}) + \dd s^2 (\mathcal{M}_{2d}) + L \dd s^2 (\mathbb{R}^{8-2d}), \nn
\Theta &=& \vol(\mathbb{R}^{1,1}) \wedge \dd (L^{-1} J_{2d}),
\eea
where $d=2, 3, 4$. In each case we require the existence of globally defined $SU(d)$ structures, specified by everywhere non-zero forms $J_{2d}, \Omega_{2d}$ on $\mathcal{M}_{2d}$. The accompanying torsion conditions follow from the $SU(4) \ltimes \mathbb{R}^8$ case of \cite{Gran:2007ps}, with the conditions for smaller structure groups being determined through decompositions of the form
\bea
J_{2d+2} &=& J_{2d} \pm e^{2d+1} \wedge e^{2d +2}, \nn
\Omega_{2d+2} &=& \Omega_{2d} \wedge (e^{2d+1} \pm i e^{2 d +2}).
\eea

As explained in detail in \cite{Gauntlett:2007ph}, the supersymmetry conditions for $AdS_3$ space-times may then be derived by introducing an $AdS_3$ radial coordinate $r$, writing the (unit radius) $AdS_3$ metric in the form
\be
\dd s^2 (AdS_3) = e^{-2 r} \dd s^2(\mathbb{R}^{1,1}) + \dd r^2,
\ee
redefining the warp factor, $L = e^{2r} \lambda$, and performing a frame rotation of the form
\be
\label{radial}
\lambda^{-\frac{1}{2}} \dd r = \sin \theta \, \hat{u} + \cos \theta \, \hat{v},
\ee
where $\theta$ parametrises the frame-rotation, which is further assumed to be independent of the $AdS_3$ radial coordinate, and $\hat{u}, \hat{v}$ are respectively unit one-forms on $\mathcal{M}_{2d}$ and the overall transverse space\footnote{For $SU(4)$ structure manifolds there is no transverse space so there $\theta = \pi/2$. }. Omitting various technicalities associated to this frame-rotation one arrives at a simple but effective derivation of the supersymmetry conditions for various $AdS_3$ space-times of type IIB supergravity. A summary of the outcome may be encapsulated in Table 1 which we reproduce from \cite{Gauntlett:2007ph}.

\begin{table}[h]
\label{tab:susy}
\begin{center}
\begin{tabular}{|c|c|c|c|}
\hline
wrapped brane & manifold & supersymmetry & R symmetry \\
\hline
K\"{a}hler 2-cycle & $CY_2$ & $\mathcal{N}=(4,4)$ & $SO(4) \times U(1)$ \\
K\"{a}hler 2-cycle & $CY_3$ & $\mathcal{N}=(2,2)$ & $U(1) \times U(1)$ \\
K\"{a}hler 2-cycle & $CY_4$ & $\mathcal{N}=(0,2)$ & $U(1)$ \\
\hline
\end{tabular}
\caption{Wrapped D3-brane geometries and their supersymmetry.}
\end{center}
\end{table}

As can be seen from the above table, in each case the cycle being wrapped is the same, but as the dimensionality of the Calabi-Yau $n$-fold ($CY_n$) increases, the preserved supersymmetry decreases. For D3-branes wrapping K\"{a}hler two-cycles in $CY_2$ manifolds, one can generically have $SO(4) \times U(1)$ R symmetry provided the radial direction (\ref{radial}) involves a rotation. Upon analytic continuation, one recovers the half-BPS LLM solutions \cite{Lin:2004nb} with isometry $\mathbb{R} \times SO(4) \times SO(4) \times U(1)$, however there appear to be no known $AdS_3$ space-times in this class. On the contrary, when $\theta =  0$, i.e. when the radial direction is purely transverse, one recovers the well known $AdS_3 \times S^3 \times CY_2$ solution\footnote{Specialising to $CY_2 = T^4$ and performing T-dualities we arrive at the usual form of the D1-D5 near-horizon sourced by three-form RR flux. We also remark that the geometry sourced by five-form flux and three-form flux are also related via fermionic T-duality \cite{Berkovits:2008ic} as explained in \cite{OColgain:2012ca}.} with R symmetry $SO(4)$. In either case the supersymmetry is $\mathcal{N} = (4,4)$.

For D3-branes wrapping K\"ahler two-cycles in $CY_3$, supersymmetry is reduced to $\mathcal{N} = (2,2)$, while the associated R symmetry group is  $U(1) \times U(1)$. Examples of these space-times can be found in the literature \cite{Maldacena:2000mw, Naka:2002jz}. Finally, for D3-branes wrapping K\"{a}hler two-cycles in $CY_4$ the dual SCFTs preserve $\mathcal{N} = (0,2)$ supersymmetry and the $U(1)$ Killing direction is dual to the R symmetry. A rich set of examples of these geometries exist in the literature \cite{Benini:2012cz, Benini:2013cda,  Gauntlett:2006ns, Gauntlett:2006qw, Naka:2002jz, Gauntlett:2006af}. In the notation of \cite{Gauntlett:2007ph}, the metric and flux may be expressed as
\bea
\label{class_sol}
\dd s^2_{10} &=& \lambda^{-1} \dd s^2(AdS_3) + \lambda \dd s^2 (\mathcal{M}_6) + \lambda^{-1} ( \dd \psi +B)^2, \\
\label{fiveform1} \Theta &=& \vol(AdS_3) \wedge \left[ \dd (\lambda^{-2} (\dd \psi + B)) -2 \lambda^{-1} J \right],
\eea
where $\partial_{\psi}$ is the Killing vector dual to the R symmetry.
The $SU(3)$ structure manifold $\mathcal{M}_6$ is subject to the the conditions \cite{Gauntlett:2007ph}:
\bea
\label{kahler} \dd J &=& 0, \\
\label{lambda} J^2 \wedge \dd B &=& \tfrac{2}{3} \lambda^2 J^3, \\
\label{omegadiff} \dd \Omega &=& 2 i ( \dd \psi + B) \wedge \Omega.
\eea
The first condition implies that $\mathcal{M}_6$ is a K\"ahler manifold, while the last condition simply identifies the Ricci form $\mathcal{R} = 2 \dd B$.


\subsection{D3-branes with $\mathcal{N} = (0,2)$ SCFTs duals}
\label{sec:D3N02}
Now that we have covered $AdS_3$ space-times arising from D3-branes wrapping K\"ahler two-cycles in Calabi-Yau manifolds in a general manner, here we focus on the particular case where the manifold is  $CY_4$. Since this case preserves the least amount of supersymmetry, it includes geometries dual to two-dimensional SCFTs with $\mathcal{N} = (2,2)$ and $\mathcal{N} = (4,4)$ supersymmetry as special cases.

While the characterisation of wrapped D3-branes \cite{Gauntlett:2007ph} presented in the previous section offers a welcome sense of overview, henceforth we switch to the notation of \cite{Gauntlett:2006ns}, which is itself based on the work of \cite{Kim:2005ez}. The generic $AdS_3$ solutions corresponding to wrapped D3-branes are then of the form \cite{Gauntlett:2006ns},
\bea
\label{gensol}
\dd s^2 &=& L^2 \left[ e^{2A} \dd s^2 (AdS_3) + \tfrac{1}{4} e^{2A} ( \dd z +P)^2 + e^{-2A} \dd s^2(\mathcal{M}_6)  \right] , \nn
\label{fiveform2} F_5 &=&  L^4 \vol(AdS_3) \wedge \left[ \tfrac{1}{2} J - \tfrac{1}{8} \dd ( e^{4A} (\dd z+ P) ) \right] \nn
&& \phantom{xxxxxxxxxxx} + \tfrac{1}{16} L^4 \left[ J \wedge \mathcal{R} \wedge (\dd z + P) + \tfrac{1}{2} *_6 \dd R \right],
\eea
where $L$ is an overall scale factor, $*_6$ refers to Hodge duality with respect to the metric of the K\"{a}hler space, $ \dd P = \mathcal{R}$, with $ \mathcal{R}$ being the Ricci form on $\mathcal{M}_6$\footnote{The Ricci form is defined by $\mathcal{R}_{ij}  = \frac{1}{2} R_{ijkl} J^{kl}$, where $R_{ijkl}$ is the Riemann tensor. Recall also that the Ricci scalar $R$ and the Ricci tensor $R_{ij}$ may be expressed in terms of the Ricci form as $R = J^{ij} \mathcal{R}_{ij}$ and $R_{ij} = - J_{i}^{~k} \mathcal{R}_{kj}$.}.
The warp factor is related to the Ricci scalar through $8 e^{-4A} = R$, a relation that can be inferred from (\ref{lambda}). The closure of $F_5$ leads to the differential condition on the curvature (\ref{diffM6}). Finally, to make direct comparison with the previous incarnation of this solution (\ref{class_sol}), one can simply redefine
\be
\lambda = e^{-2A}, \quad z = 2 \psi, \quad P = 2 B, \quad \Omega = e^{i z} \tilde{\Omega},
\ee
where we have added a tilde to differentiate between complex forms. The five-form fluxes (\ref{fiveform1}) and (\ref{fiveform2}) are related up to a factor of $-4$ and follow from the choice of normalisation adopted in \cite{Kim:2005ez}. This point should be borne in mind when making comparisons.

\subsection*{Examples}
To get better acquainted with the form of the general soution, we can consider some supersymmetric solutions that will correspond later to the vacua of our gauged supergravities. We begin with the well-known $AdS_3 \times S^3 \times T^4$ solution corresponding to the near-horizon geometry of two intersecting D3-branes. Via T-duality it is related to the D1-D5 near-horizon where the geometry is supported by a RR three-form.

To rewrite the solution in terms of the general description (\ref{gensol}), we take
\bea
A &=& 0, \nn
\dd z + P  &=& (\dd \phi_3 - \cos \phi_1 \dd \phi_2), \nn
\dd s^2(\mathcal{M}_6) &=& \dd s^2(T^4) + \tfrac{1}{4} \left( \dd \phi_1^2 + \sin^2 \phi_1 \dd \phi_2^2 \right),
\eea
where $\phi_i$ parametrise the coordinates on the $S^3$ normalised to unit radius, the same radius as the $AdS_3$ factor.  Despite this solution fitting into the general ten-dimensional framework, it preserves sixteen supercharges and is dual to a SCFT with $\mathcal{N} = (4,4)$ supersymmetry.

Before illustrating the most general solution of \cite{Benini:2012cz, Benini:2013cda} in its ten-dimensional guise, we can satisfy the required supersymmetry condition
\be
\label{susycond}
a_1 + a_2 + a_3 = - \kappa,
\ee
where $\kappa$ is the curvature of the Riemann surface $\Sigma_{\frak{g}}$, more simply through setting all the $a_i$ equal,  $a_i = \frac{1}{3}$, and taking the Riemann surface to be a unit radius Hyperbolic space, $\kappa=-1$. This solution originally featured in \cite{Naka:2002jz}. With these simplifications the solution reads
\bea
\label{constA}
\dd s^2 &=& \tfrac{4}{9} \dd s^2(AdS_3) + \tfrac{1}{3} \dd s^2(H^2) + \sum^3_{i=1} \dd \mu_i^2 + \mu_i^2 (\dd \varphi_i + \hat{A})^2, \\
F_5 &=& (1+*) \left[ - \tfrac{32}{81} \vol(AdS_3) \wedge \vol(H^2) - \tfrac{4}{27} \vol(AdS_3) \wedge \sum_{i=1}^3 \dd (\mu_i^2) \wedge ( \dd \varphi_i + \hat{A}) \right], \nonumber
\eea
where the $\mu_i$ are constrained so that $\sum_{i=1}^3 \mu_i^2 =1$.
Note now that all $A_i$ are equal, $A_i = \hat{A}$, and  $\dd \hat{A} = - \frac{1}{3} \vol(H^2)$. It is easy to determine the one-form $K = \frac{1}{2} e^{2A} (\dd z +P)$ corresponding to the R symmetry  direction
\be
K = \tfrac{2}{3} \left[ \sum_{i=1}^3 \mu_i^2 (\dd \varphi_i + \hat{A})  \right],
\ee
and check that it has the correct norm $K^2 = e^{2A} = \frac{4}{9}$ \cite{Kim:2005ez}. Taking into account the factor of $-4$ in the definitions of the flux, and also setting $L=1$, we then learn from comparing (\ref{gensol}) with (\ref{constA}) that
\be
- \tfrac{32}{81}  \vol(H^2)  - \tfrac{4}{27}  \sum_{i=1}^3 \dd (\mu_i^2) \wedge ( \dd \varphi_i + \hat{A}) = -2 J + \dd (e^{2A} K).
\ee

One can then determine $J$
\be
J = \tfrac{4}{27} \vol(H^2) + \tfrac{2}{9} \sum_{i=1}^3 \dd (\mu_i^2) \wedge  (\dd \varphi_i+\hat{A}),
\ee
which comes with the correct factor of $\vol(H^2)$,
\bea
\dd s^2(\mathcal{M}_6) &=& \tfrac{4}{27} \dd s^2 (H^2) + \tfrac{4}{9} \biggl[ \dd \mu_1^2 + \dd \mu_2^2 + \dd \mu_3^2+ \mu_1^2 \mu_2^2 (\dd \varphi_1 - \dd \varphi_2)^2  \nn &+& \mu_1^2 \mu_3^2 (\dd \varphi_1 - \dd \varphi_3)^2 + \mu_2^2 \mu_3^2 ( \dd \varphi_2 - \dd \varphi_3)^2 \biggr],
\eea
so that $\vol(\mathcal{M}_6) = \tfrac{1}{3!} J^3$. Observe also that $J$ is independent of $K$ since $ \mu_i \dd \mu_i = 0$ follows from the fact that the $\mu_i$ are constrained.  In addition, the final difference in angular coordinates $\varphi_2 - \varphi_3$ can be written as a linear combination of the other two, so we only have four directions separate from those along the $H^2$. As a further consistency check, we have confirmed that the Ricci scalar for $\mathcal{M}_6$ is $R = 8 e^{-4A}$, in line with our expectations.

We can now repeat for general $a_i$ subject to the single constraint (\ref{susycond}). This also comprises the only example we discuss where the warp factor $A$ is not a constant. In the notation of \cite{Benini:2012cz, Benini:2013cda}, the ten-dimensional solution is
\bea
\dd s^2 &=& \Delta^{\frac{1}{2}} \left[ e^{2 f} \dd s^2 (AdS_3) + e^{2g} \dd s^2(\Sigma_{\frak{g}}) \right] + \Delta^{-\frac{1}{2}} \sum_{i=1}^3 X_i^{-1} ( \dd \mu_i^2 + \mu_i^2 ( \dd \varphi_i + A^i)^2), \\
F_5 &=& (1+*) \vol(AdS_3) \wedge \sum_{i=1}^3 e^{3f+2g} \biggl[  2 X^i (X_i^2 \mu_i^2-  \Delta) \vol(\Sigma_{\frak{g}}) -  \frac{a_i}{2 \,e^{4 g} X_i^2} \dd (\mu_i^2) \wedge (\dd \varphi+ A_i)  \biggr], \nonumber
\eea
where
\be
\Delta = \sum_{i=1}^3 X_i \mu_i^2, \quad X_1 X_2 X_3 = 1,
\ee
and as before the $\mu_i$ are constrained. The constrained scalars $X_i$ can be expressed in terms of two scalars $\varphi_i$ in the following way
\be \label{x} X_1 = e^{-\frac{1}{2} \left(
\frac{2}{\sqrt{6}} \varphi_1 + \sqrt{2} \varphi_2 \right) }, \quad
X_2 = e^{-\frac{1}{2} \left( \frac{2}{\sqrt{6}} \varphi_1 - \sqrt{2}
\varphi_2 \right) }, \quad X_3 = e^{ \frac{2}{\sqrt{6}} \varphi_1 }.
\ee

To give the full form of the solution one also needs to specify the values of the various warp factors $e^{f}, e^{g}$ and scalars $X_i$  \cite{Benini:2012cz}\footnote{The solutions with $\frak{g} =1$ were studied in \cite{Almuhairi:2011ws}, while for $\frak{g}=0$, $\frak{g}>1$, modulo issues related to the range of the parameters, the solutions can be mapped to (4.6) of \cite{Cucu:2003bm} through interchanging the scalars $\phi_1 \leftrightarrow -\phi_2$ and redefining the parameters accordingly $a_i = -\epsilon m_i/({m_1 + m_2 +m_3})$, where $\epsilon = 1$ for $\Sigma_{\frak{g}} = S^2$ and $\epsilon =-1$ for $\Sigma_{\frak{g}} = H^2$.}:
\bea
e^{f} &=& \frac{2}{X_1 + X_2 + X_3}, \quad ~~~~~~~~~ e^{2g} = \frac{a_1 X_2 + a_2 X_1}{2}, \nn
X_1 X_3^{-1} &=& \frac{a_1}{a_3} \frac{(a_2+a_3-a_1)}{(a_1 + a_2 - a_3)}, \quad
X_2 X_3^{-1} =  \frac{a_2}{a_3} \frac{(a_1+a_3-a_2)}{(a_1 + a_2 - a_3)}.
\eea

From the higher-dimensional perspective afforded to us here, the canonical R symmetry corresponds with the Killing vector \cite{Benini:2013cda}
\be
\label{Rvec}
\partial_{\psi} = 2 \sum_{i=1}^3 \frac{X_i}{X_1 + X_2 + X_3} \partial_{\varphi_i}.
\ee
Again, one is in a position to determine the dual one-form
\be
K = e^{f} \Delta^{-\frac{1}{2}} \sum_{i=1}^3 \mu_i^2 ( \dd \varphi_i + A_i),
\ee
and confirm that it squares correctly $K^2 = e^{2A} = \Delta^{\frac{1}{2}} e^{2 f}$. Proceeding in the same fashion as above, one can then determine $J$
\be
J =  \sum_{i=1}^3 \tfrac{1}{4} \left[ - \frac{\Theta}{a_i (2 a_i + \kappa)} e^{3 f} \dd (\mu_i^2) \wedge (\dd \varphi_i + A_i) + 2 a_i  (2 a_i + \kappa) \frac{\Theta}{\Pi} \mu_i^2 e^{3f} \vol(\Sigma_{\frak{g}}) \right],
\ee
where we have adopted the notation of \cite{Benini:2013cda}, namely
\bea
\label{thetapi}
\Theta &=& a_1^2 + a_2^2 + a_3^2 - 2 (a_1 a_2 + a_1 a_3 + a_2 a_3), \nn
\Pi &=& (-a_1 + a_2 + a_3) (a_1 -a_2 +a_3) (a_1 + a_2 -a_3).
\eea
The accompanying expression for the manifold $\mathcal{M}_6$ is
\bea
\dd s^2(\mathcal{M}_6) &=& \Delta e^{2g + 2 f} \dd s^2(\Sigma_{\frak{g}}) + e^{2 f} \biggl[ X_1^{-1} \dd \mu_1^2 + X_2^{-1} \dd \mu_2^2 + X_3^{-1} \dd \mu_3^2 \nn &+& \frac{X_3}{\Delta} \mu_1^2 \mu_2^2 (X_2 \DD \varphi_1 - X_1 \DD \varphi_2)^2 + \frac{X_2}{\Delta} \mu_1^2 \mu_3^2 (X_3 \DD \varphi_1 - X_1 \DD \varphi_3)^2\nn &+& \frac{X_1}{\Delta} \mu_2^2 \mu_3^2 (X_3 \DD \varphi_2 - X_2 \DD \varphi_3)^2 \biggr],  \eea
where we have further defined $\DD \varphi_i = \dd \varphi_i + A_i$.
One can check it is consistent with the expression for $J$ and furthermore that one recovers the previous expressions upon simplification, i.e. setting $a_i = \frac{1}{3}$, $\kappa=-1$.

These solutions will all be utilised later when we come to discuss three-dimensional gauged supergravities with vacua corresponding to the above supersymmetric solutions. In the next section, we begin by discussing an example of a generic reduction, in other words one where the warp factor is not a constant, by providing further details of the reduction and resulting  three-dimensional $\mathcal{N} =2$ supergravity initially reported in \cite{Karndumri:2013iqa}.

\section{An example of a generic reduction}
\label{sec:generic}
In this section we illustrate an example of a generic reduction, where we use the word ``generic" to draw a line between dimensional reductions with non-trivial warp factors from the ten-dimensional perspective, and those that are direct products. Recall that, in addition to the famous KK reductions based on spheres \cite{s51, dewit, s41},  which give rise to maximal gauged supergravities in lower dimensions, generic KK reductions based on gaugings of R symmetry groups, notably gaugings of $U(1)$ R symmetry \cite{Gauntlett:2006ai, Gauntlett:2007ma} and $SU(2)$ R symmetry \cite{Gauntlett:2007sm,Jeong:2013ifc} exist despite the internal space not being a sphere. This observation leads to the natural conjecture \cite{Gauntlett:2007ma} that gaugings of R symmetry groups are intimately connected to the existence of consistent KK dimensional reductions. Here should be no exception, so we expect that one can gauge the existing $U(1)$ R symmetry present in (\ref{gensol}) and reduce to three dimensions.

However, in contrast to similar reductions to four and five dimensions, for instance \cite{Gauntlett:2006ai, Gauntlett:2007ma}, here in addition to retaining the gauge field from the R symmetry gauging, we also require an additional scalar so that the three-dimensional gauged supergravity fits into the structure of $\mathcal{N} =2$ gauged supergravity as laid out in \cite{deWit:2003ja}. More concretely, we require an even number of scalars to constitute a K\"{a}hler scalar manifold. While the reduction we discuss presently assumes additional structure for the $\mathcal{M}_6$, i.e. the existence of a Riemann surface, it would be interesting to identify truly generic reductions without having to specify the internal six-dimensional K\"{a}hler manifold.

Here we will present further details of the dimensional reduction from five-dimensional $U(1)^3$  gauged supergravity to three-dimensional $\mathcal{N} =2$ gauged supergravity reported in \cite{Karndumri:2013iqa}. While not being the most general reduction, from the ten-dimensional vantage point it provides a neat example of a reduction where the warp factor, and the associated Ricci scalar of the internal $\mathcal{M}_6$, is not a constant. We also do not need to address the full embedding of the three-dimensional theory in ten dimensions, since we can work with the $U(1)^3$ gauged supergravity in five dimensions.

The bosonic sector of the action for five-dimensional $U(1)^3$ gauged supergravity can
be found in \cite{Cvetic:1999xp}. It arises as a consistent
reduction from type IIB on $S^5$, so it is directly connected to ten dimensions\footnote{The bosonic sector also appears as a reduction from $D=11$ supergravity \cite{mention} where it is based on the existence of near-horizon black holes \cite{Fareghbal:2008eh}. Interestingly, one can start from $D=11$ and reduce to $D=4$ $U(1)^4$ gauged supergravity, which, for consistency, requires $F^i \wedge F^j = 0$. Taking a near-horizon limit prescribed in \cite{Fareghbal:2008eh} one finds the bosonic sector of $D=5$ $U(1)^3$ gauged supergravity, without such a condition.} via the equations of motion, and corresponds to the special case where only the
$SO(2)^3$ Cartan subgroup of $SO(6)$ is gauged.
The action reads
\bea
\label{U13act}
\mathcal{L}_5 &=& R * \mathbf{1} - \tfrac{1}{2}
\sum_{i=1}^2 \dd \varphi_i \wedge * \dd \varphi_i  - \tfrac{1}{2} \sum_{i=1}^3
X_i^{-2} F^i \wedge * F^i \nn &+& 4 g^2 \sum_{i=1}^3 X_i^{-1} \vol_5 +
F^1 \wedge F^2 \wedge A^3,
\eea
where $g$ is the gauge coupling and the constrained scalars $X_i$ we have defined earlier (\ref{x}). From varying the potential with respect to the scalars it
is easy to see that there is only a single supersymmetric $AdS_5$
vacuum at $X_i=1$.

As commented in \cite{Cvetic:1999xp}, or by inspection from the equations of motion in appendix \ref{sec:U13}, one can consistently truncate the theory by setting first $\varphi_2 = 0$ implying that $X_1 = X_2 = X_3^{-1/2}$. This truncation is consistent provided $F^1 = F^2$. Furthermore, one can take an additional step and set $\varphi_1 = 0$ leading to minimal gauged supergravity in five dimensions.

\subsection*{Dimensional reduction}
As it turns out, this dimensional reduction can be performed consistently at the level of the action. Simply put, this means that we can adopt the space-time metric Ansatz
\be
\dd s^2_5 = e^{-4C} \dd s^2_3 + e^{2C} \dd s^2(\Sigma_{\frak{g}})
\ee
where $\Sigma_{\frak{g}}$ is a constant curvature Riemann surface of genus $\frak{g}$ and we have used $C$ to denote the scalar warp factor in five dimensions.  In addition, we have orchestrated the warp factors so that we arrive directly in Einstein frame in three dimensions.

The metric on the Riemann surface may be
expressed as \be \dd s^2(\Sigma_{\frak{g}}) = e^{2 h} \left( \dd x^2 + \dd y^2\right),
\ee
where the function $h$ depends on the curvature $\kappa$ of the Riemann
surface. It is respectively, $ h = - \log \left( (1+ x^2 + y^2)/2
\right)$ ($\kappa=1$), $h = \log (2 \pi)/2$ ($\kappa=0$) and $h = -
\log(y)$ ($\kappa=-1$), depending on whether the genus is $\frak{g} =0, \frak{g}=1$, or $\frak{g}>1$.
In addition, one takes the following Ansatz for the field strengths,
\be
\label{U13gauge}
F^i = G^i - a_i \vol(\Sigma_{\frak{g}}),
\ee
where closure of $F^i$ ensures that $a_i$ are constants and $G^i$ is closed, $G^i = \dd B^i$.

In doing the reduction at the level of the action the following expression for the five-dimensional Ricci scalar is useful
\be
R * \mathbf{1} = {R} *_3 \mathbf{1} - 6 \dd C \wedge *_3 \dd C + 2 \kappa e^{-6C} *_3 \mathbf{1}.
\ee

The resulting three-dimensional action in Einstein frame is
\bea \label{Einsteinact}
\mathcal{L}^{(3)} &=&  R *_3 \mathbf{1} - 6 \dd C \wedge *_3 \dd C -
\tfrac{1}{2} \sum_{i=1}^2 \dd \varphi_i \wedge *_3 \dd \varphi_i  -
\tfrac{1}{2} e^{4C} \sum_{i=1}^3 X_i^{-2}  G^i \wedge *_3 G^i \nn &+&
\left( \sum_i^3 \left[ 4 g^2 e^{-4C} X_i^{-1} - \tfrac{1}{2} e^{-8C}
a_i^2 X_i^{-2} \right]  + 2 \kappa e^{-6C} \right) * \mathbf{1}+
\mathcal{L}^{(3)}_{\textrm{top}}, \eea
where the topological term takes the form
\be
\mathcal{L}^{(3)}_{\textrm{top}} = a_1 B^2 \wedge G^3 + a_2 B^3 \wedge G^1
+ a_3 B^1 \wedge G^2.
\ee

We remark that the reduction and the resulting potential appeared previously in \cite{Cucu:2003yk}. In appendix \ref{sec:U13}, we have confirmed that it is indeed consistent.

\subsection*{Dualising the action}
Now that we have the action, we would like to rewrite it in the form of a three-dimensional non-linear sigma model coupled to supergravity so that we can make contact with three-dimensional gauged supergravities in the literature \cite{deWit:2003ja}. We take our first steps in that direction by dualising the gauge fields, or more appropriately, their field strengths, and replacing them with scalars:
\bea
\label{duality} G^1 &=& X^{2}_1 e^{-4C}
* \DD Y_1, \quad \DD Y_1= \dd Y_1 + a_3 B^2 +a_2 B^3,\nn
G^2 &=& X^{2}_2
e^{-4C} * \DD Y_2, \quad \DD Y_2= \dd Y_2 + a_1 B^3 +a_3 B^1 ,\nn
G^3 &=&
X^{2}_3 e^{-4C} * \DD Y_3, \quad \DD Y_3= \dd Y_3 + a_1 B^2 +a_2 B^1.
\eea
Through these redefinitions, we can recast the action (\ref{Einsteinact}) in the following form
 \bea \label{dualact} \mathcal{L}^{(3)} &=&  R *
\mathbf{1} - 6 \dd C \wedge * \dd C - \tfrac{1}{2} \sum_{i=1}^2 \dd \varphi_i
\wedge * \dd \varphi_i  - \tfrac{1}{2} e^{-4C} \sum_{i=1}^3 X_i^{2}  \DD Y_i
\wedge * \DD Y_i \nn &+&  \mathcal{L}^{(3)}_{\textrm{pot}} + a_1 B^2 \wedge
G^3 + a_2 B^3 \wedge G^1  + a_3 B^1 \wedge G^2, \eea
where we have omitted the explicit form of the potential as it will play no immediate role. We have also dropped all subscripts for Hodge duals on the understanding that we are now confining our interest to three dimensions. Note that the
Chern-Simons terms are untouched and when we
vary with respect to $B^i$ we recover the duality conditions (\ref{duality}), so it should be clear that the equations of motion are the same and we have just rewritten the action.

At this point, before blindly stumbling on, we will attempt to motivate the expected gauged supergravity. Firstly, we know from the Killing spinor analysis in \cite{Benini:2013cda} that the $AdS_3$ solutions generically preserve four supersymmetries, meaning we are dealing with $\mathcal{N} =2$ supersymmetry in three dimensions. Indeed, for $\mathcal{N} =2$, we have precisely an $SO(2)$ R symmetry group under which the gravitini transform and in this case the target space is a K\"{a}hler manifold with the scalars pairing into complex conjugates. Naturally, a prerequisite for a K\"{a}hler manifold is that we have an even number of scalars, and we observe that after dualising, this is indeed the case. So, we will now push ahead and identify some features of the $\mathcal{N} =2$ gauged supergravity.

To identify the scalar manifold it is good to diagonalise the scalars by redefining them in the following way
 \bea W_1 &=& 2 C + \tfrac{1}{\sqrt{6}} \varphi_1 +
\tfrac{1}{\sqrt{2}} \varphi_2, \nn W_2 &=& 2 C + \tfrac{1}{\sqrt{6}}
\varphi_1 - \tfrac{1}{\sqrt{2}} \varphi_2, \nn W_3 &=& 2 C -
\tfrac{2}{\sqrt{6}} \varphi_1. \eea In terms of the original $X_i$
these new scalars are simply $e^{W_i} = e^{2C} X_i^{-1}$.

With these redefinitions, the K\"ahler manifold now assumes the simple form
\be \mathcal{L}^{(3)}_{\textrm{scalar}} = -
\tfrac{1}{2} \sum_{i=1}^3 \left[ \dd W_i \wedge * \dd W_i + e^{-2 W_i} \DD Y_i
\wedge * \DD Y_i \right] \ee
and we are in a position to identify it as $[SU(1,1)/U(1) ]^3$. The K\"ahler structure of the scalar target space can be made fully explicit through the introduction of the K\"ahler potential of the form
\be
\mathcal{K} = - \sum_{i=1}^3 \log (\Re z_i),
\ee
where we have introduced complex coordinates $z_i = e^{W_i} + i Y_i$. This means that the metric for the manifold is  $g_{i \bar{i}} = \partial_i \partial_{\bar{i}} \mathcal{K} = \frac{1}{4} e^{-2 W_i}$, where $\partial_{i} = \partial_{z_i}, \partial_{\bar{i}} = \partial_{\bar{z}_i}$.

Having identified the scalar manifold and the K\"{a}hler potential, we turn our attention to the scalar potential. In the language of three-dimensional $\mathcal{N} =2$ gauged supergravity \cite{deWit:2003ja},  the scalar potential is comprised of two components, a $T$ tensor and a superpotential $W$:
\be
\mathcal{L}^{(3)}_{\textrm{pot}} =  8 T^2 - 8 g^{i \bar{i}}
\partial_i T \partial_{\bar{i}} T + 8 e^{\mathcal{K}} |W|^2 - 2 g^{i
\bar{i}} e^{\mathcal{K}} D_i W D_{\bar{i}} \bar{W}, \ee
where the
K\"{a}hler covariant derivative is $D_i W \equiv \partial_i W +
\partial_i \mathcal{K} W$ and $W$ is holomorphic, so $\partial_i \bar{W} = \partial_{\bar{i}} W = 0$.  While $W$ plays a natural role when eleven-dimensional supergravity is reduced on $S^2 \times CY_3$ to three dimensions \cite{Colgain:2010rg}, whenever the R symmetry is gauged, consistency demands that $W=0$. Thus, to make contact with the literature, we face the simpler task of identifying the correct $T$ tensor and making sure that the potential is recovered.

After rewriting the scalars, the potential takes the more symmetric form
\bea \label{pot} \mathcal{L}^{(3)}_{\textrm{pot}} &=& 4 g^2 \left[
e^{-W_1-W_3} + e^{-W_2  -W_3} + e^{-W_1-W_2} \right]  +  2 \kappa
e^{-W_1 -W_2 -W_3} \nn &-& \tfrac{1}{2} \left[ a_1^2 \,
e^{-2(W_2+W_3)} + a_2^2 \, e^{-2(W_1+W_3)} + a_3^2 \,
e^{-2(W_1+W_2)} \right]. \eea
Note that in performing the reduction we have not been picky about supersymmetry and \textit{a priori}, neglecting the gauge coupling $g$, which can be set to one, the constants $\kappa$ and $a_i$ are unrelated. However, setting $g=1$ for simplicity, one can find the appropriate $T$ tensor
 \bea T = -\tfrac{1}{4}\left[ a_1 e^{-W_2 -W_3} + a_2 e^{-W_1 -
W_3} +a_3 e^{-W_1 - W_2} \right] + \tfrac{1}{2} \left[ e^{-W_1}
+e^{-W_2} + e^{-W_3}\right], \label{T} \eea
and check that it reproduces the potential on the nose provided (\ref{susycond}) is satisfied.
This is precisely the condition identified in \cite{Benini:2012cz, Benini:2013cda} for supersymmetry to be preserved. Though it happens that the existence of what is commonly referred to as a ``superpotential", in this case $T$, could conceivably be related to some fake supersymmetry structure for the theory, the fact that we recover the supersymmetry condition is reassuring. In fact, in appendix \ref{sec:KSE} we reduce some of the Killing spinor equations and show that they also lead to the same $T$ tensor. Thus, once the potential (and also $T$) is extremised, the Killing spinor equations are satisfied.

\subsection*{Central charge and exact R symmetry}
At this stage it should be obvious that we have a potential with a supersymmetric critical point provided condition (\ref{susycond}) holds. Furthermore, once we extremise $T$, we in turn extremise the potential and arrive at the supersymmetric $AdS_3$ vacuum. As discussed in \cite{Karndumri:2013iqa}, the extremization of the $T$ tensor offers a natural supergravity counterpart for $c$-extremization \cite{Benini:2012cz, Benini:2013cda}.  Recall that $c$-extremization has been proposed for SCFTs with $\mathcal{N} = (0,2)$ supersymmetry as a means to identify the exact central charge and R symmetry where ambiguities exist due to the $U(1)$ R symmetry mixing with other global $U(1)$ symmetries that may be present.

Like the trial $c$-function proposed in \cite{Benini:2012cz, Benini:2013cda}, $T$ is also quadratic and comes from squaring the moment maps $\mathcal{V}^i$
\be
\label{T}
T =2 \mathcal{V}^i \Theta_{ij} \mathcal{V}^{j},
\ee
contracted with the embedding tensor $\Theta_{ij}$ \cite{deWit:2003ja}, where the index $i$ ranges over the various $U(1)$ symmetries, which for the immediate example, $i=1, 2, 3$. In addition, since the embedding tensor also appears in the Chern-Simons terms in the action, it also related to the 't Hooft anomaly coefficients which appear in the trial $c$-function for $c$-extremization \cite{Benini:2012cz, Benini:2013cda}. Indeed, for the class of wrapped D3-brane geometries discussed in \cite{Benini:2012cz, Benini:2013cda} this can all be made precise through the relations \cite{Karndumri:2013iqa}
\be
\label{exactCR}
c_R = 3 \eta_{\Sigma} d_G T^{-1}, \quad R = 2 \, \mathcal{V}^i T^{-1} Q_i,
\ee
where $c_R$ is the exact central charge, $R$ is the exact R symmetry, $\eta_{\Sigma}$ is related to the volume of the Riemann surface, $\eta_{\Sigma} = 2 \pi \vol(\Sigma_{\frak{g}})$, $d_G$ is the dimension of the gauge group and $Q_i$ denotes the charges corresponding to the $U(1)$ currents.

All that remains to do is simply to identify the minimum of the potential by extremising $T$. The critical point of $T$ corresponds to the following values for the scalars:
\begin{eqnarray}
W_1&=&\ln \left[\frac{a_2a_3}{a_2+a_3-a_1}\right],\qquad W_2=\ln
\left[\frac{a_1a_3}{a_1+a_3-a_2}\right]\nonumber \\
W_3&=&\ln
\left[\frac{a_1a_2}{a_1+a_2-a_3}\right].\label{critical_point}
\end{eqnarray}
Once written in terms of $C$, $\varphi_1$ and $\varphi_2$ or in
terms of $C$ and $X_i$, this precisely gives the $AdS_3$ critical
point of \cite{Benini:2012cz}. Then, slotting the critical value of $T$ into the (\ref{exactCR}), we arrive at the exact central charge and R symmetry,
\bea
\label{c} c_R &=& - 12 \eta_{\Sigma} N^2 \frac{ a_1 a_2 a_3 }{\Theta}, \\
\label{r} R &=& \frac{2 a_i ( 2 a_i + \kappa)}{\Theta},
\eea
where we have made use of (\ref{thetapi}) to display the result. In deriving (\ref{c}) we have used the fact that the dimension of the gauge group at large $N$ is $d_G = N^2$, while for (\ref{r}) it is good to use the fact that the moment map is $\mathcal{V}_i = \frac{1}{4} e^{-W_i}$. The central charge and R symmetry agree with those quoted  in \cite{Benini:2012cz, Benini:2013cda} and reproduce the coefficients of the Killing vector corresponding to the R symmetry (\ref{Rvec}).

\section{Less generic reductions}
Experience suggests that it is much easier to construct KK reduction Ans\"atze for direct product solutions than those that are warped products. This should come as no surprise since warped products are often more involved and consequently it may not be easy to identify a symmetry principle to guide the construction of a fitting Ansatz. For dimensional reductions from ten or eleven dimensions to five-dimensional gauged supergravities admitting $AdS_5$ vacua, the restrictions are quite clear. Starting with coset reductions \cite{s51, Cassani:2010na, Bena:2010pr}, through generic Sasaki-Einstein reductions \cite{Cassani:2010uw, Gauntlett:2010vu, Liu:2010sa, Skenderis:2010vz} to the more general cases, the richness of the reduced theory gradually decreases until one is left with minimal gauged supergravity \cite{Gauntlett:2006ai, Gauntlett:2007ma}. For warped $AdS_5$ solutions, only reductions to minimal gauged supergravity are known, with a notable exception being KK reductions \cite{OColgain:2011ng} based on $Y^{p,q}$ spaces \cite{Gauntlett:2004zh, Gauntlett:2004yd}, which when uplifted to eleven dimensions, the vacua correspond to warped solutions.

In this section we will discuss KK reductions to three dimensions confined to the special case where the K\"{a}hler manifold is a product of K\"ahler-Einstein spaces. As a direct consequence, (\ref{diffM6}) simplifies to
\be \label{algM6} R^2 = 2 R_{ij} R^{ij}.
\ee
A nice treatment of this special case can be found in \cite{Gauntlett:2006ns} which we follow.  We take the internal K\"{a}hler manifold to be a product of a set of two-dimensional K\"{a}hler-Einstein metrics
\be
\dd s^2 (\mathcal{M}_6) = \sum_{i=1}^{3} \dd s^2 (KE_2^{(i)}).
\ee
Since $\mathcal{M}_6$  now has constant curvature, it is easy to satisfy (\ref{algM6}). The Ricci form for $\mathcal{M}_6$ takes the form
\be
\label{RKE}
\mathcal{R} = \sum_{i=1}^3 l_i J_i,
\ee
where $J_i$ are the K\"{a}hler forms of the constituent metrics and the constants $l_i$ are zero, positive or negative depending on whether the metric is locally that on $T^2$, $S^2$ or $H^2$. We also have the one-form connection $P = \sum_i P_i$ with $\dd P= \sum_i l_i J_i$. Slotting (\ref{RKE}) into (\ref{algM6}) we find a single constraint on the $l_i$
\be
\label{cond1}
l_1 l_2 + l_1 l_3 + l_2 l_3 = 0,
\ee
and discover that the overall warp factor is determined,
\be
\label{warp}
e^{-4 A} = \tfrac{1}{8} R = \tfrac{1}{4} \sum_i l_i .
\ee
Finally, the expression for the five-form flux (\ref{gensol}) simplifies and assumes the following form
\be
F_5 = (1+ *) L^4 \vol(AdS_3) \wedge \frac{1}{2 \sum_i l_i} \left[J_1 (l_2 + l_3) + J_2 (l_1 + l_3) + J_3 (l_1 + l_2) \right].
\ee

We now can make some comments. Demanding that the ten-dimensional space-time has the correct signature, we require $R > 0$ from (\ref{warp}). In the light of (\ref{cond1}), this means that the potential solutions are constrained to be either $S^2 \times T^4$ or $S^2 \times S^2 \times H^2$. The first option here corresponds to the famous intersecting D3-branes solution, while the second case was considered in \cite{Naka:2002jz}.
We note that when the $KE_2^{(i)}$ space is $H^2$, it is a well-known fact that one can quotient the space without breaking supersymmetry leading to a compact Riemann surface with genus $\frak{g} > 1$. The Ricci tensor for these solutions can be found in appendix \ref{sec:KE6ricci}.

\subsection{Twists of SCFTs with Sasaki-Einstein duals }
\label{sec:N1twist}
In this section we will discuss KK reductions on the first class of products of K\"{a}hler-Einstein spaces by confining our attention to spaces with curvature, $l_i \neq 0$. For simplicity, we will take $l_1 = l_2$, and the requirement that the scalar curvature of the internal $\mathcal{M}_6$ be positive (\ref{warp}) subject to (\ref{cond1}) means that there is only one case, namely $\mathcal{M}_6 = H^2 \times KE_4$, where $KE_4$ is a positively curved K\"{a}hler-Einstein manifold\footnote{Suitable choices for $KE_4$ include $S^2 \times S^2$, $\mathbb{C} P^2$ and del Pezzo $\dd P_k$, $k=3,\dots, 8$.}. For concreteness, we take $(l_1, l_2, l_3) = (2, 2, -1)$ so that the $H^2$ is canonically normalised.

Our next task is to construct a ten-dimensional Ansatz. While we could begin from scratch, we can incorporate some results from the literature as, in the end, a natural question concerns how they may be related. So we opt to kill two birds with one stone by simply reducing the IIB reduction on a generic Sasaki-Einstein five-manifold $SE_5$ \cite{Cassani:2010uw, Gauntlett:2010vu, Skenderis:2010vz, Liu:2010sa} further to three dimensions on a constant curvature Riemann surface ($H^2$). We will follow the notation of \cite{Gauntlett:2010vu} and subsequent comments are in the context of that work.

To achieve our goal, we make two simplifications. Firstly, we truncate out the complex two-form $L_2$, since as our internal space is now six-dimensional, a complex $(2,0)$-form, $\Omega_2$, is less natural. We can easily replace it with a field coupling to the complex $(3,0)$-form $\Omega_3$ via the five-form flux, but this will simply give us an additional complex scalar.
More importantly, one can ask what is the fate of the complex scalars $\xi$ and $\chi$ under dimensional reduction. Recall that they feature prominently in embeddings of holographic superconductors \cite{Gubser:2009qm} (see also \cite{Gauntlett:2009dn, Gauntlett:2009bh}). However, since $\xi, \chi$ couple to the graviphoton $A_1$, it is not possible to twist $A_1$ in the usual way to produce a supersymmetric $AdS_3$ vacuum without truncating out $\xi$ and $\chi$. As such, we will have nothing to say about models for holographic superconductivity here. Moreover, as the same fields support the non-supersymmetric Romans' vacuum in five dimensions, we do not expect to find an analogue in three dimensions that follows from the reduction procedure.

The five-dimensional action in Einstein frame can be found in (3.10) of \cite{Gauntlett:2010vu}. With the above simplifications taken onboard,  for completeness, we reproduce the kinetic term
\begin{eqnarray} \label{kineticEinframeSE}
{\cal L}_{\textrm{kin}}^{(5)} &=& R \ \textrm{vol}_5 -\tfrac{28}{3} \dd U \wedge * \dd U  -\tfrac{8}{3}  \dd U \wedge * \dd V - \tfrac{4}{3} \dd V \wedge * \dd V -\ft12 e^{2\phi} \dd a \wedge * \dd a \nonumber \\ && -\ft12 \dd \phi \wedge * \dd \phi
-2 e^{-8U} K_1 \wedge * K_1 - e^{-4U-\phi} H_1 \wedge * H_1 - e^{-4U+\phi} G_1 \wedge * G_1\nn
&& -\ft12 e^{\frac{8}{3}(U+V)} F_2\wedge * F_2
 - e^{-\frac{4}{3}(U+V)} K_2 \wedge *K_2
 -\ft{1}{2} e^{\frac{4}{3}(2U-V)-\phi} H_2 \wedge * H_2
 \\ && -\ft{1}{2} e^{\frac{4}{3}(2U-V)+\phi} G_2 \wedge * G_2
 -\ft{1}{2} e^{\frac{4}{3}(4U+V)-\phi} H_3 \wedge * H_3 -\ft{1}{2} e^{\frac{4}{3}(4U+V)+\phi} G_3 \wedge * G_3, \nonumber
 \end{eqnarray}
the scalar potential
\begin{eqnarray} \label{potentialEinframe}
{\cal L}_{\textrm{pot}}^{(5)} &=& \Big[ 24e^{-\frac{2}{3}(7U+V)} -4e^{\frac{4}{3}(-5U+V)} -8e^{-\frac{8}{3}(4U+V)} \Big] \textrm{vol}_5 \; ,
  \end{eqnarray}
and the topological terms are given by the expression
\begin{eqnarray}\label{ltopmmm}
{\cal L}_{\textrm{top}}^{(5)}&=& -A_1\wedge K_2\wedge K_2 -(\dd k -2E_1 -2
A_1) \wedge [\dd B_2 \wedge (\dd c-2C_1) + (\dd b-2B_1) \wedge \dd C_2] \nonumber \\ &&
+A_1 \wedge (\dd k  -2E_1) \wedge [(\dd b-2B_1) \wedge \dd C_1 -\dd B_1 \wedge (\dd c-2C_1)] \nonumber \\ && +2A_1 \wedge dE_1 \wedge (\dd b-2B_1) \wedge (\dd c-2C_1) +
A_1 \wedge (\dd b-2B_1) \wedge (\dd c-2C_1) \wedge F_2 \nonumber \\ &&   -4
C_2 \wedge \dd B_2 .
\end{eqnarray}
In turn, the above fields can be written in terms of various potentials and scalars in five dimensions
\bea
G_1 &=& \dd c - 2 C_1 - a \dd b +2 a B_1, \nn
H_1 &=& \dd b - 2 B_1, \nn
K_1 &=& \dd k -2 E_1 -2 A_1, \nn
F_2 &=& \dd A_1, \nn
G_2 &=& \dd C_1 - a \dd B_1, \nn
H_2 &=& \dd B_1, \nn
K_2 &=& \dd E_1 + \tfrac{1}{2} (\dd b -2 B_1) \wedge (\dd c - 2 C_1),
\eea
thus ensuring the that ten-dimensional Bianchi identities (appendix \ref{sec:IIBsugra}) for the fluxes hold. In total we have 7 scalars $U, V, k, b, c$ including the axion $a$ and dilaton $\phi$, 4 one-form potentials $A_1, B_1, C_1, E_1$ and 2 two-form potentials $B_2, C_2$.

\subsection*{Dimensional reduction}
Having introduced the five-dimensional theory, we are in a position to push ahead with the same reduction as section \ref{sec:generic} to three dimensions on a constant curvature Riemann surface $\Sigma_{\frak{g}}$.
We consider the usual metric Ansatz\footnote{Here $C$ without subscript will denote the scalar warp factor and $C_1$ is a one-form.}
\be
\dd s^2_5 = e^{-4C} \dd s^2_3 + e^{2C} \dd s^2(\Sigma_{\frak{g}}),
\ee
where warp factors have been chosen so that we end up in Einstein frame, and for the moment, we will assume that we have a constant curvature Riemann surface and not specify its curvature $\kappa$. Supersymmetry will later dictate that $\kappa < 0$. As for the rest of the fields, the five-dimensional scalars reduce to three-dimensional scalars. The fact that the field strengths $H_1, G_1$ appear in the Einstein equation mean that we cannot twist with respect to $B_1$ and $C_1$ since such a twisting is inconsistent with the assumption that the Riemann surface is constantly curved. This leaves $A_1$ and $E_1$, or their field strengths, which we twist in the following way
\bea
\label{twisting}
K_2 &=& - \e \vol (\Sigma_{\frak{g}}) + \tilde{K}_2, \nn
F_2 &=& \e \vol (\Sigma_{\frak{g}}) + \tilde{F}_2,
\eea
where tildes denote three-dimensional field strengths. $\epsilon$ is dictated to be a constant through $ F_2 = \dd A_1$ and no twisting along $K_1$ imposes the requirement that we twist $K_2$ in the opposite way. This latter point is also in line with our expectation that one can further truncate the theory to minimal gauged supergravity through $K_1 = 0, K_2 = - F_2$ in five dimensions \cite{Gauntlett:2010vu}.

Since we are not twisting $B_1, C_1$, the  field strengths $G_1, H_1, G_2, H_2$ reduce directly to three dimensions. On the contrary, we can consider a decomposition for the three-form field strengths $G_3, H_3$ on the condition that we respect the symmetries of $\Sigma_{\frak{g}}$. So we can decompose
\be
C_2 = e \vol(\Sigma_{\frak{g}}) + \tilde{C}_2, \quad B_2 = f \vol(\Sigma_{\frak{g}}) + \tilde{B}_2,
\ee
leading to two new scalars $e, f$ in the process. The corresponding field strengths can then be written as
\bea
G_3 &=&  M_1 \wedge \vol (\Sigma_{\frak{g}}) + g \vol_3, \quad M_1 =  \dd e - a \dd f + \tfrac{1}{2} \epsilon ( \dd c - 2 C_1 - a \dd b +2 a B_1) ,\nn
H_3 &=& N_1 \wedge \vol (\Sigma_{\frak{g}}) +  h \vol_3,   \quad N_1 = \dd f + \tfrac{1}{2} \epsilon (\dd b - 2 B_1).
\eea
One can check that this choice is consistent with the closure of the Bianchi identities.

The scalars $g, h$ are, up to an integration constants $\lambda_1, \lambda_2$, set by the equations of motion
\bea
g &=& - 4 e^{-\frac{4}{3}(4 U +V) - \phi - 8C} (\lambda_1 + f) \nn
h &=&   4 e^{-\frac{4}{3}(4 U +V) + \phi - 8C}  (\lambda_2 + e - a (\lambda_1 + f)).
\eea
We will normalise these so that $\lambda_i = 1$.

We now reduce directly at the level of the action and take care to check in appendix \ref{sec:H2KE4} that one gets the same result from reducing the equations of motion, thus guaranteeing the consistency of the reduction. Dropping tildes, as only the three-dimensional fields remain, the resulting kinetic terms are
\bea
\label{N1twist}
\mathcal{L}^{(3)}_{\textrm{kin}} &=& R \vol_3 - 6 \dd C \wedge * \dd C - \tfrac{28}{3} \dd U \wedge * \dd U - \tfrac{8}{3} \dd U \wedge * \dd V- \tfrac{4}{3} \dd V \wedge * \dd V
- \tfrac{1}{2} e^{2 \phi} \dd a \wedge * \dd a \nn &-& \tfrac{1}{2} \dd \phi \wedge * \dd \phi - 2 e^{-8U} K_1 \wedge * K_1  - e^{-4 U + \phi} G_1 \wedge * G_1 -   e^{-4 U - \phi} H_1 \wedge * H_1
\nn &-& \tfrac{1}{2} e^{\frac{4}{3} (4 U + V) + \phi -4 C} M_1 \wedge * M_1 - \tfrac{1}{2} e^{\frac{4}{3} (4 U + V) - \phi -4 C} N_1 \wedge * N_1- e^{-\frac{4}{3} (U + V) + 4 C} K_2 \wedge * {K}_2 \nn &-&  \tfrac{1}{2} e^{\frac{8}{3} (U+V)+4 C} {F}_2 \wedge * {F}_2 - \tfrac{1}{2} e^{\frac{4}{3}(2 U -V) + \phi + 4C} G_2 \wedge * G_2 - \tfrac{1}{2} e^{\frac{4}{3} (2 U - V) - \phi + 4 C} H_2 \wedge * H_2, \nn
\eea
while those of the scalar potential take the form
\bea
\mathcal{L}^{(3)}_{\textrm{pot}} &=& e^{-4 C} \biggl[ 2 \kappa e^{-2C}  + 24 e^{-\frac{2}{3} (7 U + V)} - 4 e^{\frac{4}{3} (-5 U +V)} - 8 e^{-\frac{8}{3} (4 U + V)}  \nn &-& \tfrac{1}{2} \e^2 e^{-4C} \left( e^{\frac{8}{3} (U + V)} + 2  e^{-\frac{4}{3} (U + V) } \right) - {8 e^{-\frac{4}{3} (4 U + V) - \phi - 4C } ( 1+ f)^2 }  \nn &-&  { 8 e^{-\frac{4}{3} (4 U + V) + \phi - 4C } ( 1+ e - a (1 +  f) )^2  } \biggr] \vol_3.
\eea
The topological term is then given by the expression
\bea
\mathcal{L}^{(3)}_{\textrm{top}} &=& 2 \e  {A}_1 \wedge {K}_2 - { 4 (1 + e) A_1 \wedge \dd B_1 + 4 (1 +f) A_1 \wedge \dd C_1}  \nn
&-& \epsilon E_1 \wedge K_2
+ 2 E_1 \wedge \left[ \dd f \wedge (\dd c -2 C_1 ) - \dd e \wedge (\dd b - 2 B_1) + \tfrac{3}{4} \epsilon (\dd b - 2 B_1) \wedge (\dd c -2 C_1)\right] \nn
&+&{2 k \left[ (\dd f + \tfrac{1}{2} \epsilon (\dd b - 2 B_1) ) \wedge \dd C_1 - (\dd e + \tfrac{1}{2} \epsilon (\dd c -2 C_1) )\wedge \dd B_1 \right] }.
\eea

Now is an opportune time to identify the supersymmetric $AdS_3$ vacuum. This can be done by comparing directly with (6.9) of \cite{Gauntlett:2006ns} (see also \cite{Gauntlett:2006qw}). For concreteness we can take $KE_4 = S^2 \times S^2$ to exhibit the explicit solution, but one can consider other choices. The form of the space-time metric before rescaling is
\bea
\label{explicitsol}
\dd s^2 &=& L^2 \biggl[  \tfrac{2}{\sqrt{3}} \dd s^2(AdS_3) + \tfrac{\sqrt{3}}{2} \left( \frac{\dd x^2 + \dd y^2}{y^2} \right)  +   \tfrac{\sqrt{3}}{2} \sum_{i=1}^2 \tfrac{1}{2} (\dd \theta_i^2 + \sin^2 \theta_i \dd \phi_i^2)  \nn && \phantom{xxxxxxxxxxxxxxxxxxxx} \tfrac{1}{2 \sqrt{3}} \left(\dd z - \frac{\dd x}{y} - \sum_i \cos \theta_i \dd \phi_i  \right)^2 \biggr],
\eea
where $AdS_3$ is normalised to unit radius and all normalisations for the $H^2$, parametrised by $(x,y)$, and two $S^2$'s, parametrised by $(\theta_i,  \phi_i)$ are now explicit. We have also reintroduced an overall scale factor $L$. We omit the five-form flux as it will not provide any new information and it is enough to compare the ten-dimensional metrics.

To make meaningful comparison with the KK reduction Ansatz of \cite{Gauntlett:2010vu}, we need to compare with the following space-time Ansatz
\bea
\dd s^2 &=& e^{-\frac{2}{3} (4 U + V)} \left[ e^{-4C} \dd s^2_3 + e^{2C} \dd s^2(\Sigma_{\frak{g}}) \right] + e^{2U} \dd s^2(KE_4) + e^{2V} (\eta + A_1)^2,
\eea
where $\dd \eta = 2 J$ and the K\"ahler-Einstein metric $g_{ij}$ with positive curvature is normalised so that $R_{ij} = 6 g_{ij}$. To make the connection, we first rescale the $KE_4$ factor in (\ref{explicitsol}) by a factor of three, take $L^2 = 2/(3 \sqrt{3})$ and make the following identifications
\bea
\label{eta}
(\eta + A_1) = \tfrac{1}{3} \left( \dd z -\cos \theta_1 \dd \phi_1 - \cos \theta_2 \dd \phi_2 - \frac{\dd x}{y} \right).
\eea
The supersymmetric $AdS_3$ vacuum can then be identified
\be
U = V = 0, \quad C = - \tfrac{1}{2} \log 3, \quad e = f = -1,
\ee
where $\kappa = -1$, since the $H^2$ was normalised to unit radius, and $\epsilon = - \frac{1}{3}$ follows from (\ref{eta}). One can indeed check that this choice leads to a critical point of the potential and that the $AdS_3$ radius of the three-dimensional space-time is $\ell = \frac{2}{9}$.

\subsection*{Further truncation \& supergravity}
In this subsection we consider the above action with the three-form fluxes truncated out by setting $b = c = B_1 = C_1 = B_2 = C_2 = 0$, $e=f = -1$.  Even from the ten-dimensional perspective, it is known that it is always consistent to perform this truncation to just the metric, fields in the five-form flux and the axion and dilaton\footnote{In performing this truncation we remove the six scalars coming from the RR and NS three-form fluxes. In general, it is possible to see that one always has an $SU(1,1)/U(1)$ factor, but it is not clear if the remaining twelve scalars constitute a K\"ahler manifold. It is also possible that the vacuum spontaneously breaks $\mathcal{N}=4$ supersymmetry to $\mathcal{N}=2$, for example \cite{Gauntlett:2010vu} in five dimensions. We leave this point to future work.}.

We now recast the simpler action in the more familiar language of three-dimensional gauged supergravity. In part this will involve dualising the one-form potentials. To do so we redefine the following fields 
\bea
{K}_2 &=&  e^{\frac{4}{3} (U + V) - 4 C} * \DD Y_2, \quad \DD Y_2 = \dd Y_2 + \tilde{B}_2 \nn
{F}_2 &=& e^{-\frac{8}{3} (U+V)-4 C} * \DD Y_3, \quad \DD Y_3 = \dd Y_3 + \tilde{B}_3,
\eea
while, at the same time, adding the following additional CS terms
\be
\delta \mathcal{L}^{(3)}_{\textrm{top}} = 2 \tilde{B}_2 \wedge K_2 + \tilde{B}_3 \wedge F_2.
\ee
The covariant derivatives are chosen so that the equations of motion are still satisfied once $\tilde{B}_i$ are integrated out. We can then redefine $K_1$
\be
K_1 = \tfrac{1}{2} \DD Y_1, \quad \DD Y_1 =  \left( \dd Y_1 - 4 E_1 - 4 A_1 \right),
\ee
and finally introduce the following scalars
\be
W_1 = -4 U, \quad  W_2 =  \tfrac{2}{3} (U+V) -2 C , \quad W_3 =-\tfrac{4}{3} (U+V)-2 C.
\ee
The scalar manifold is now $[SU(1,1)/U(1)]^4$, which should be familiar from previous analysis, and the kinetic term for the action becomes
\bea
\mathcal{L}_{\textrm{kin}} &=& - \tfrac{1}{2}  \dd W_1 \wedge * \dd W_1 - \tfrac{1}{2} e^{2 W_1} \DD Y_1 \wedge * \DD Y_1  -  \dd W_2 \wedge * \dd W_2 - e^{2 W_2} \DD Y_2 \wedge * \DD Y_2\nn  &-& \tfrac{1}{2}  \dd W_3 \wedge * \dd W_3 - \tfrac{1}{2} e^{2 W_3} \DD Y_3 \wedge * \DD Y_3
 - \tfrac{1}{2} \dd \phi \wedge * \dd \phi - \tfrac{1}{2} e^{2 \phi} \dd a \wedge * \dd a.
\eea

We can thus introduce the complex coordinates
\be
z_i = e^{-W_i} + i Y_i,~~i = 1, 2, 3,  \quad
z_4 = e^{-\phi} + i a,
\ee
allowing us explicitly to write the K\"{a}hler potential $\mathcal{K}$ as
\be
\mathcal{K} = -  \log (\Re z_1) - 2 \log (\Re z_2) - \log (\Re z_3) - \log (\Re z_4).
\ee
While we could have made this point earlier, it is now clear that the axion $a$ and the dilaton $\phi$ decouple completely and can be truncated out. They also do not feature in the scalar potential.

In terms of the other scalars the potential takes the form
\bea
\mathcal{L}_{\textrm{pot}} &=&  \biggl[ 2 \kappa e^{2 W_2 + W_3}  + 24 e^{W_1+W_2+W_3}- 4 e^{2(W_1+W_2)} - 8 e^{2(W_1+W_3)}  \nn &-& \tfrac{1}{2} \e^2 \left( e^{4 W_2} + 2  e^{2(W_2 +W_3)} \right) \biggr] \vol_3.
\eea
We can then work out the corresponding $T$ tensor in terms of $\epsilon$ and $\kappa$,
\bea
T &=& -\frac{\e}{4} e^{2 W_2} - \frac{\e}{2} e^{W_2 + W_3} - e^{W_1+W_2} + e^{W_1+W_3} - \frac{\kappa}{2 \epsilon} e^{W_3}.
\eea
We note that $\kappa$ and $\epsilon$ are not independent and we require $\kappa = 3 \epsilon$ so that the $T$ tensor reproduces the potential. Once they are identified in this way, and taking into account the fact that $\kappa < 0, \epsilon < 0$, one finds a vacuum at
\be
W_1 = 0, ~~W_2 = W_3 = -\log (-\epsilon) \Rightarrow U = V = 0, ~~C =  \tfrac{1}{2} \log \left( -\epsilon \right).
\ee
Setting $\epsilon = - \frac{1}{3}$, we arrive at the result quoted previously.

\subsection*{Central charge and R symmetry}
In fact we have already discussed the central charge for this case as it corresponds to a particular example in section \ref{sec:generic}, namely $a_i = \frac{1}{3}, \kappa = -1$, thus ensuring that (\ref{susycond}) is satisfied. However, to avoid the onerous task of rescaling metrics and comparing solutions, we can simply recalculate the central charge using the standard holographic prescription \cite{Brown:1986nw, Henningson:1998gx}
\be
c_R = \frac{3 \ell}{\phantom{x}2 G^{(3)}},
\ee
where $\ell$ is the $AdS_3$ radius and $G^{(3)}$ the three-dimensional Newton's constant. Using the conventions of \cite{Benini:2012cz,Benini:2013cda} where $G^{(3)} = 1/( 4 \eta_{\Sigma} N^2)$, one can check that the result agrees with (\ref{c}) when $a_i = \frac{1}{3}$.

It is also of interest here to ask about the R symmetry? The ten-dimensional origin of our reduction makes it clear that there is only a single $U(1)$ R symmetry, so there is no ambiguity. However, without this insight, we can ask what the three-dimensional theory can tell us about the R symmetry. Once we truncate out $K_1$, we have essentially two $U(1)$ symmetries and the moment maps $\mathcal{V}^i$ associated to these can be worked out by comparing the $T$ tensor (\ref{T}) with the CS term in the action. We find that the components of the embedding tensor are $\Theta_{23} = 2 \epsilon, \Theta_{22} = - 2 \epsilon$ and, for agreement, the moment maps are
\be
\mathcal{V}^2 = \tfrac{1}{4} e^{W_2}, \quad \mathcal{V}^3 = -\tfrac{1}{4} e^{W_3},
\ee
where $i = 2, 3$ label the $U(1)$'s associated to the gauge fields $E_1$ and $A_1$ respectively. We can then extract the R symmetry
\be
R = - \tfrac{2}{3} U(1)_2 + \tfrac{2}{3} U(1)_3 ,
\ee
where we have again used indices to distinguish the $U(1)$'s. We can now compare to our earlier result (\ref{r}) by inserting $a_i = \frac{1}{3}$ and one arrives at the same numbers, up to a relative sign. This relative sign can be traced to the relative sign in (\ref{twisting}) and by simply changing the sign of $A_1$ in ten dimensions, one can find perfect agreement.

\subsection{Intersecting D3-branes}
\label{sec:D3}
In this section we discuss dimensional reductions to three dimensions for intersecting D3-branes. Some of the work presented here will not be new and will recover the recent work of \cite{Detournay:2012dz}. Although we could approach this task directly from a ten-dimensional Ansatz, it is handier to make use of an intermediate reduction to six dimensions on a Calabi-Yau two-fold \cite{Duff:1998cr}, details of which can be found in appendix \ref{sec:CYred}.

As such, we adopt the same strategy as \cite{Detournay:2012dz}, but an important distinction is that we will not impose truncations directly in six dimensions and then reduce. Instead, we will reduce directly so that we can unify the reductions presented in \cite{Detournay:2012dz}. In addition, we will make statements about the underlying gauged supergravity, an aspect that was overlooked in \cite{Detournay:2012dz}. Note that it is expected that the three-dimensional gauged supergravity be a theory with $\mathcal{N} =4$ supersymmetry, so that the scalar manifold is a product of quaternionic manifolds \cite{deWit:2003ja}, but this falls outside of our scope here and we hope to address this question in future work. Finally, we remark that these reductions are related to those of \cite{Colgain:2010rg} via T-duality and uplift, a point that is fleshed out in appendix \ref{sec:3dreds}.

So the task now is to perform the reduction on $S^3$, written as a Hopf-fibration, from the six-dimensional theory presented in \cite{Duff:1998cr} to extract a three-dimensional gauged supergravity.  Strictly speaking we are then doing a reduction on the D1-D5 near-horizon or its S-dual F1-NS5, so further T-dualities along $CY_2  = T^2 \times T^2$ will be required to recover the intersecting D3-brane vacuum discussed previously. We will come to this point in due course.

\subsection*{Dimensional reduction}
Starting from the six-dimensional theory in appendix \ref{sec:CYred}, we adopt the natural space-time Ansatz
\be
\label{6dmet}
\dd s^2_6 = e^{-4U - 2V} \dd s^2_3 + \tfrac{1}{4} e^{2U} \dd s^2(S^2) + \tfrac{1}{4} e^{2V} ( \dd z +P + A_1),
\ee
where $U, V$ are warp factors and $A_1$ is a one-form with legs on the three-dimensional space-time. Our Ansatz fits into the overarching description for supersymmetric $AdS_3$ solutions from wrapped D3-branes presented earlier with choice $(l_1, l_2, l_3) = (0, 0, 4)$. In contrast to \cite{Detournay:2012dz}, this means that $P = - \cos \theta \dd \phi$ so that $\dd P = \vol(S^2) = 4 J_3$. In addition, $A =0$ follows from (\ref{warp}).

For the three-form fluxes, we consider the following Ansatz
\bea
\label{6dflux}
F_3 &=& G_0 \tfrac{1}{2} ( \dd z + P + A_1) \wedge J_3 + G_1 \wedge J_3 + G_2 \wedge \tfrac{1}{2} (\dd z + P + A_1) + g e^{-6 U -3 V} \vol_3 \\
H_3 &=& \sin \alpha  ( \dd z + P + A_1) \wedge J_3 + H_1 \wedge J_3 + H_2 \wedge \tfrac{1}{2} (\dd z + P + A_1) + h e^{-6U - 3 V} \vol_3, \nonumber
\eea
where the Bianchi identities (see appendix \ref{sec:IIBsugra}) determine the following:
\bea
G_ 0 &=& 2 \left( \cos \alpha - \sin \alpha \chi_1 \right), \nn
G_1 &=& \dd c - \chi_1 \dd b - 2 (C_1 - \chi_1 B_1) - (\cos \alpha - \sin \alpha \chi_1) A_1, \nn
G_2 &=& \dd C_1 - \chi_1 \dd B_1, \nn
H_1 &=& \dd b - 2 B_1 - \sin \alpha A_1, \nn
H_2 &=& \dd B_1.
\eea
Here $\chi_1$ is the scalar axion of type IIB supergravity and we have introduced the constant $\alpha$, scalars $b, c$ and one-form potentials $B_1, C_1$. The remaining scalars, $\phi_i, \chi_i$ $=1, 2$, of the six-dimensional theory simply descend to become three-dimensional scalars.

We now plug our Ansatz into the equations of motion of the six-dimensional theory (\ref{6deom1}) - (\ref{6deom7}), the details of which can be found in appendix \ref{sec:CYred}. In the process one determines the form for $g, h$:
\bea
\label{g} g &=& 2 e^{-\phi_1+\phi_2-V -2 U} (\cos \alpha + \sin \alpha \chi_2), \\
\label{h} h &=& 2 e^{\phi_1 + \phi_2-V - 2 U} [ \sin \alpha - \cos \alpha \chi_2 + (\cos \alpha + \sin \alpha \chi_2) \chi_1],
\eea
where we have normalised the integration constants for later convenience.

One finds that the equations of motion all come from varying the following three-dimensional action:
\be
\mathcal{L}^{(3)} = \mathcal{L}^{(3)}_{\textrm{kin}} + \mathcal{L}^{(3)}_{\textrm{pot}} + \mathcal{L}^{(3)}_{\textrm{top}},
\ee
where the kinetic term is
\bea
\label{D3act}
\mathcal{L}^{(3)}_{\textrm{kin}} &=& R \vol_3 - \tfrac{1}{2} \dd \phi_1 \wedge * \dd  \phi_1 - \tfrac{1}{2} e^{2 \phi_1} \dd \chi_1 \wedge * \dd \chi_1 - \tfrac{1}{2} \dd \phi_2 \wedge * \dd  \phi_2 \nn &-& \tfrac{1}{2} e^{2 \phi_2} \dd \chi_2 \wedge * \dd \chi_2
- 6 \dd U \wedge * \dd U - 4 \dd U \wedge * \dd V - 2 \dd V \wedge * \dd V \nn
&-& \tfrac{1}{2} e^{-\phi_1- \phi_2 -4 U } H_1 \wedge * H_1 - \tfrac{1}{2} e^{\phi_1- \phi_2 -4 U } G_1 \wedge * G_1 - \tfrac{1}{2} e^{-\phi_1- \phi_2 +4 U } H_2 \wedge * H_2 \nn &-& \tfrac{1}{2} e^{\phi_1- \phi_2 +4 U } G_2 \wedge * G_2 - \tfrac{1}{8} e^{4 U + 4 V} F_2 \wedge * F_2,
\eea
and the scalar potential takes the form
\bea
\mathcal{L}^{(3)}_{\textrm{pot}} &=& \biggl[ 8 e^{-6U-2V}  - 2 e^{-8 U} -2 e^{\phi_1 + \phi_2 -8 U - 4V} \left[ \sin \alpha- \cos \alpha \chi_2 + (\cos \alpha + \sin \alpha \chi_2) \chi_1 \right]^2 \nn &-& 2 e^{-\phi_1 + \phi_2 -8 U - 4 V} ( \cos \alpha + \sin \alpha \chi_2)^2 -  2 e^{-\phi_1 -\phi_2 - 8 U -4 V} \sin^2 \alpha \nn &-&  2 e^{\phi_1 -\phi_2 - 8 U -4 V} (\cos \alpha - \sin \alpha \chi_1)^2 \biggr] \vol_3.
\eea
Finally, the topological term takes the simple form
\bea
\mathcal{L}^{(3)}_{\textrm{top}} &=& \chi_2 \left( H_1 \wedge G_2 - G_1 \wedge H_2 \right)-  (\cos \alpha C_1 + \sin \alpha B_1) \wedge F_2.
\eea

When $ U = V = \phi_i = \chi_i  = c = b = A_1 = B_1 = C_1 = 0$, the above scalar potential has a critical point corresponding to either the D1-D5 near-horizon, its S-dual, or a one parameter interpolating vacuum. We have chosen the integration constants so that an $SL(2, \mathbb{R})$ transformation, parametrised by the constant $\alpha$,
\be
\left( \begin{array}{c} C_2 \\ B_2 \end{array} \right) \rightarrow \left( \begin{array}{cc} \cos \alpha & - \sin \alpha \\ \sin \alpha & \cos \alpha \end{array} \right) \left( \begin{array}{c} C_2 \\ B_2 \end{array} \right)
\ee
 takes one from the vacuum supported by a RR three-form flux ($\alpha = 0$) to the vacuum supported by a NS three-form flux ($\alpha = \frac{\pi}{2}$). In each case the $AdS_3$ radius is unity. It is known more generally that the effect of an $SL(2, \mathbb{R})$ transformation is simply to rotate the Killing spinors \cite{Ortin:1994su}\footnote{In this immediate context, see \cite{OColgain:2012ca}.}, so supersymmetry is unaffected.

\subsection*{Ten-dimensional picture}
As we have reached our three-dimensional theory through the result of two steps, a reduction on a Calabi-Yau two-fold \cite{Duff:1998cr} and a further reduction generalising the recent work of \cite{Detournay:2012dz}, here we wish to pause to consider the higher-dimensional picture. We would also like to recast the KK reduction Ansatz in terms of the generic form of wrapped D3-branes. Specialising to $CY_2 = T^2 \times T^2$, we can perform two T-dualities along the second $T^2$ leading to the following NS sector with the metric in string frame:
\bea
\label{10dsol1}
\dd s^2_{10} &=& e^{\frac{1}{2} (\phi_1 + \phi_2)} \left[ e^{-4 U -2 V} \dd s^2_3 + \tfrac{1}{4} e^{2 U} \dd s^2(S^2) + \tfrac{1}{4} e^{2 V} (\dd z + P + A_1)^2 \right] \\
&& \phantom{xxxxxxxx} + e^{\frac{1}{2} (\phi_1 - \phi_2)} \dd s^2 (T^2_1) + e^{ -\frac{1}{2} (\phi_1 - \phi_2)} \dd s^2(T^2_2), \nn
H_3 &=& \left[ 2 \sin \alpha J_3  + H_2 \right] \wedge \tfrac{1}{2} ( \dd z + P + A_1) + H_1 \wedge J_3 + h e^{-6U - 3 V} \vol_3, \\
\tilde{\phi} &=& \tfrac{1}{2} ( \phi_1 + \phi_2),
\eea
where $\tilde{\phi}$ is the new ten-dimensional dilaton.
Note that the three-form flux $H_3$ is not affected by the T-duality. The accompanying RR fluxes then take the form
\bea
\label{10dsol2}
F_5 &=& \left[ G_0J_2 \wedge J_3 + g e^{\phi_1 - \phi_2 + 2 U + V} J_1 \wedge J_3  \right] \wedge \tfrac{1}{2} (\dd z+ P + A_1)
+ e^{\phi_1 - \phi_2 +4 U} * G_2 \wedge J_1 \wedge J_3 \nn &+&  G_1 \wedge J_2 \wedge J_3
+ \left[G_2 \wedge J_2 -e^{\phi_1 - \phi_2-4 U} * G_1 \wedge J_1 \right] \wedge  \tfrac{1}{2} (\dd z + P + A_1) \nn
&+& G_0 e^{\phi_1 - \phi_2-8 U - 4 V} \vol_3 \wedge J_1 + g e^{-6 U -3 V} \vol_3 \wedge J_2, \nn
F_3 &=& \dd \chi_1 \wedge J_2 - \dd \chi_2 \wedge J_1,
\eea
where $J_1 = \vol(T^2_1), J_2 = \vol(T^2_2)$ and, as before, $J_3 = \frac{1}{4} \vol(S^2)$ and there is no axion, $F_1 = 0$.

\subsection*{Further truncations}
Even if we dualise the gauge fields in the action (\ref{D3act}), since we have an odd number of scalars and $\mathcal{N} =2$ supergravity in three dimensions has a K\"{a}hler scalar manifold, one will need to truncate out some fields to find a gauged supergravity description. In this subsection we consider some further truncations and make contact with the work of \cite{Detournay:2012dz} in the process.

Setting $\alpha = \chi_i =  c = A_1 = C_1 =  0$, $\phi_1 = \phi_2 = \phi$, $U = -V$,  and finally employing the following identification
\bea
B_1 = \hat{A}
\eea
one can check that our action can be brought to the form of (4.7) of \cite{Detournay:2012dz}:
\bea
\label{GDact1}
\mathcal{L}^{(3)} &=& R \vol_3 + (4 e^{-4U} -2 e^{-8U})\vol_3- \dd \phi \wedge * \dd  \phi
- 4 \dd U \wedge * \dd U \nn
&-& \tfrac{1}{2} e^{-2 \phi-4 U } H_1 \wedge * H_1 - \tfrac{1}{2} e^{-2 \phi +4 U } H_2 \wedge * H_2.
\eea
Note we have set $\ell = 1$ for simplicity, but this can be reinstated if one rescales the radius of the Hopf-fibre $S^3$ correctly. We have also retained the scalar field $b$, which one is required to set to zero to make direct connection with \cite{Detournay:2012dz}.

The reduction of \cite{Detournay:2012dz}, where the six-dimensional space-time is supported solely by RR flux, involves setting $\phi_1 = -\phi_2 = \phi$, $\chi_i = b =  \alpha  = B_1 = 0$. Making the further identifications
\bea
C_1 = -\hat{A}, \quad A_1 = 2A,
\eea
one arrives at
\bea
\label{GDact2}
\mathcal{L}^{(3)} &=& R \vol_3 - \dd \phi \wedge * \dd  \phi
- 6 \dd U \wedge * \dd U - 4 \dd U \wedge * \dd V - 2 \dd V \wedge * \dd V \nn
&-& \tfrac{1}{2}  e^{2 \phi -4 U } (\dd c + 2( \hat{A} - A) )  \wedge * (\dd c + 2(\hat{A} - A))  - \tfrac{1}{2} e^{2 \phi+4 U } \hat{F} \wedge * \hat{F} - \tfrac{1}{2} e^{4 U + 4 V} F \wedge * F  \nn &+&  \left[8 e^{-6 U-2 V} - 2 e^{-8U} -2 e^{-2 \phi-8U-4V} -2 e^{2 \phi-8 U -4V} \right] \vol_3 + 2 \hat{A} \wedge F.
\eea
Once one sets $c=0$ one can again confirm this is the same action as (4.17) of \cite{Detournay:2012dz}. A further truncation of action ($\phi=0 = A, U= - V)$ permits warped black string solutions,  the holographic interpretation of which was considered in \cite{Guica:2013jza}\footnote{It is easier to start with the action in \cite{Guica:2013jza} and use the EOM for $\hat{A}$ to find the form for the action above. } .

An obvious truncation not discussed in \cite{Detournay:2012dz} is the truncation to just the NS sector. In some sense this may be regarded as the S-dual of the truncation we have just discussed. We can do this by setting $\alpha = \frac{\pi}{2}$, $\chi_i = c = C_1 = 0$ and $\phi_1 = \phi_2 = \tilde{\phi}$. The resulting action is
\bea
\label{Sdualact}
\mathcal{L}^{(3)} &=& R \vol_3 -  \dd \tilde{\phi} \wedge * \dd  \tilde{\phi}
- 6 \dd U \wedge * \dd U - 4 \dd U \wedge * \dd V - 2 \dd V \wedge * \dd V \nn
&-& \tfrac{1}{2} e^{-2 \tilde{\phi} -4 U } H_1 \wedge * H_1 - \tfrac{1}{2} e^{-2 \tilde{\phi} +4 U } H_2 \wedge * H_2 - \tfrac{1}{8} e^{4 U + 4 V} F_2 \wedge * F_2 \nn
&+& \left[8 e^{-6 U-2 V} - 2 e^{-8U} -2 e^{-2 \phi-8U-4V} -2 e^{2 \phi-8 U -4V} \right] \vol_3 - B_1 \wedge F_2.
\eea
Up to a rewriting, $b = c, A_1 = 2 A, B_1 =  - \hat{A}, \tilde{\phi} = -\phi$, this action is identical to (\ref{GDact2}).

\subsection*{Rewriting the supergravity}
Here we identify the underlying gauged supergravities. As a warm-up we consider the action (\ref{GDact1}), but make a conversion from the three-dimensional Yang-Mills (YM) Lagrangian to a Chern-Simons Lagrangian following general prescriptions given in \cite{deWit:2003ja} (see also \cite{ Nicolai:2003bp, Berg:2002es}). This procedure replaces every  YM gauge field with two gauge fields and a new scalar field. This
allows us to trade the following Yang-Mills term in the action
\be
\mathcal{L}^{(3)}_{\textrm{YM}} = - \tfrac{1}{2} e^{-2\phi+4U}H_2\wedge * H_2\label{YM1}
\ee
with the terms
\be
\mathcal{L}^{(3)}_{\textrm{CS}} = - \tfrac{1}{2} e^{2 \phi - 4 U} \DD \tilde{\phi} \wedge * \DD \tilde{\phi} + H_2 \wedge \tilde{B}_1,
\ee
where $\DD \tilde{\phi} = \dd \tilde{\phi} - \tilde{B}_1$ and we now have two gauge fields $B_1, \tilde{B}_1$ and an additional scalar $\tilde{\phi}$. Varying with respect to $\tilde{B}_1$ we get
\be
\label{H2phi}
H_2+e^{2\phi-4U}*{\DD}\tilde{\phi}=0,
\ee
which, on choosing the gauge $\tilde{\phi} =0$, we can integrate out $\tilde{B}_1$ to recover the original Lagrangian. The equation of motion following from varying $B_1$ now reads
\be
\dd \tilde{B}_1 + 2 e^{-2 \phi -4 U} * H_1 = 0,
\ee
which can be shown to be equivalent to that of the original Lagrangian once one imposes (\ref{H2phi}). The equation of motion for $\tilde{\phi}$ is trivially satisfied through (\ref{H2phi}).

With these changes, the scalar kinetic term of the full Lagrangian (\ref{GDact1}) is given by
\begin{equation}
\mathcal{L}^{(3)}_{\textrm{kin}}=-\dd \phi \wedge * \dd \phi -4 \dd U\wedge * \dd U-\tfrac{1}{2} e^{-2\phi-4U} H_1 \wedge *H_1-\tfrac{1}{2}e^{2\phi-4U} {\DD}\tilde{\phi}\wedge *{\DD}\tilde{\phi}
\end{equation}
where as before $H_1=\dd b-2B_1$.
We redefine all of the scalars through
\begin{equation}
Y_1=\tilde{\phi},\qquad Y_2=b,\qquad W_1=\phi-2U,\qquad W_2=-\phi-2 U,
\end{equation}
so that the scalar kinetic term becomes
\begin{equation}
\mathcal{L}^{(3)}_{\textrm{kin}}=-\tfrac{1}{2} \sum_{i=1}^2 \left[ \dd W_i \wedge* \dd W_i+ e^{2W_i} {\DD}Y_i\wedge * {\DD}Y_i \right].
\end{equation}
The corresponding scalar manifold is clearly $[SU(1,1)/U(1)]^2$ and the K\"ahler potential is $\mathcal{K} = - \sum_i \log (\Re z_i)$, where $z_i = e^{-W_i} + i Y_i$. In terms of $W_i$, the scalar potential becomes
\begin{equation}
\mathcal{L}^{(3)}_{\textrm{pot}} = \left[ 4e^{W_1+W_2}-2e^{2(W_1+W_2)} \right]  \vol_3 \, .
\end{equation}
The corresponding $T$ tensor is found to be
\begin{equation}
T=\tfrac{1}{2}\left(e^{W_1}+e^{W_2}-e^{W_1+W_2}\right)
\end{equation}
with only one critical point at $W_1=W_2=0$.  Here it is not immediately obvious that this is the only option. Recall that for $\mathcal{N} =2$ gauged supergravity, when the R symmetry is gauged, no holomorphic superpotential can appear \cite{deWit:2003ja}. Now when the R symmetry is not gauged, as is the case here, one can consider replacing the $T$ tensor with the free energy $F = -T \pm e^{\mathcal{K}/2} W$.  However, since $e^{\mathcal{K}} = e^{W_1 + W_2}$, we can see that a problem arises with $W$ being holomorphic, so this does not appear to be an option.

We now move onto the second action that results from truncating out all the NS three-form flux fields. Referring to (\ref{10dsol1}), (\ref{10dsol2}), this means that we set $\alpha = b = B_1 = \chi_i = 0$. With this simplification, one further observes that it is consistent to set $\phi_1 = - \phi_2 = \phi$.  This is simply (\ref{GDact2}) with the scalar $c$ reinstated and $A_1$ and $C_1$ rewritten accordingly, $A_1 = 2 A$, $C_1 = - \hat{A}$.

We can now diagonalise the scalars by redefining them
\bea
W_1 &=& -\phi-2 U, \quad W_2 = \phi-2 U, \quad W_3 = -2 U -2 V,
\eea
leading to canonically normalised kinetic terms:
\bea
\mathcal{L}^{(3)}_{\textrm{kin}} = - \tfrac{1}{2} \sum_{i=1}^3 \left[ \dd W_i \wedge * \dd W_i + e^{2W_i} \DD Y_i \wedge * \DD Y_i \right]
\eea
In the process we have redefined $Y_2 = c$ so that $\DD Y_2 = \dd Y_2 + 2 (\hat{A} -A)$ and in addition dualised the one-form potentials, $A, \hat{A}$ so that
\bea
\hat{F} &=& e^{-2 \phi-4 U} * \DD Y_1, \quad \DD Y_1 = \dd Y_1 + B_1, \nn
F &=& e^{-4 U - 4 V} * \DD Y_3, \quad \DD Y_3 = \dd Y_3 + B_3.
\eea
As should be customary at this stage, we have to add a corresponding CS term so the new topological term is
\be
\mathcal{L}^{(3)}_{\textrm{top}} = 2 \hat{A} \wedge F + B_1 \wedge \hat{F} + B_3 \wedge F.
\ee

Introducing complex coordinates in the usual fashion, $z_i = e^{-W_i} + i Y_i$, $i=1, 2, 3$, the K\"{a}hler potential for the scalar manifold is $
\mathcal{K} = - \sum_{i} \log (\Re z_i)$.

In terms of our new scalars $W_i$, the potential takes a simple form and is symmetric in all the scalars $W_i$:
\be
\mathcal{L}^{(3)}_{\textrm{pot}} = 2 \left[ 4 e^{W_1 + W_2 + W_3} - e^{2(W_1+W_3)} - e^{2(W_1+W_2)} -  e^{2(W_2+W_3)}  \right] \vol_3.
\ee
A suitable choice for the corresponding $T$ tensor is
\bea
T = - e^{W_2} + \tfrac{1}{2} ( e^{W_1 +W_2} - e^{W_1 +W_3} + e^{W_2 +W_3}),
\eea
though symmetry dictates that there are other choices and we can send $W_1 \rightarrow W_2 \rightarrow W_3 \rightarrow W_1$ to uncover the other options. Regardless of how we choose $T$, the critical point is located at $W_i= 0$. Since the R symmetry is gauged, we do not expect a holomorphic superpotential.

\section{Null-warped $AdS_3$ solutions}
\label{sec:Sch}
Recently, it has been noted that null-warped $AdS_3$ solutions, or equivalently geometries exhibiting Schr\"{o}dinger symmetry with $z=2$, can be found in three-dimensional theories that arise as consistent reductions based on the D1-D5 (or its S-dual) near-horizon geometries of type IIB supergravity \cite{Detournay:2012dz}. In section \ref{sec:D3}, we identified the relevant theories in the gauged supergravity literature and here we will discuss some of the solutions. Prior to \cite{Detournay:2012dz}, it was noted that non-relativistic geometries with dynamical exponent $z=4$ could be found in an $\mathcal{N}=2$ gauged supergravity that is the consistent KK reduction of eleven-dimensional supergravity on $S^2 \times CY_3$ \cite{Colgain:2010rg}\footnote{These were mistakenly labelled null-warped $AdS_3$, but this label should be reserved solely for the $z=2$ case in the literature.}. We will now address a natural question by scanning the other gauged supergravities we have identified for non-relativistic solutions with dynamical exponent $z$.

Before doing so, we recall some facts about Schr\"{o}dinger solutions in three dimensions. Starting from an $AdS_3$ vacuum, solutions with dynamical exponent $z$ arise as solutions to Chern-Simons theories where the relevant equation is
\be
\label{CSeq}
\dd *_3 F + \frac{\kappa}{\ell} F =0,
\ee
with $F = \dd A$ and $\ell $ denotes the $AdS_3$ radius. Taking the derivative of (\ref{CSeq}), we see that $\kappa$ must be a constant. Adopting the usual form of the space-time Ansatz
\be
\dd s^2 = \ell^2 \left( - \lambda^2 r^z \dd u^2 + 2 r \dd u \dd v + \frac{\dd r^2}{4 r^2}\right),
\ee
the Einstein equation, through the components of the Ricci tensor\footnote{We have used the dreibein $e^{+} = \ell \, r^{\frac{1}{2}} \dd u,
e^{-} = \ell \, r^{\frac{1}{2}} \left( \dd v - \frac{1}{2} \lambda^2 r^{z-1} \dd u \right),
e^{r} = \ell \, \frac{\dd r}{2 r}. $} :
\be
R_{+-} = - \frac{2}{\ell^2}, \quad
R_{++} =  \frac{\lambda^2}{\ell^2} 2 z (z-1) r^{z-1}, \quad
R_{--} = 0,
\ee
determines the constant $\kappa$ in terms of the dynamical exponent, $\kappa = z$. Observe here that $\lambda$ is an arbitrary constant that can either be set to unity through rescaling the metric, or when set to zero, one recovers the unwarped $AdS_3$ vacuum.

Now the task of searching for new solutions becomes a very accessible goal; one simply has to identify $\ell$ and compare the equations of motion of the theory with (\ref{CSeq}) to extract $\kappa$ and thus $z$. For the gauged supergravity discussed in section \ref{sec:generic}, namely the theory given by the action (\ref{Einsteinact}), the $AdS_3$ radius is
\be
\ell = \frac{1}{2 T} =  -\frac{2 a_1 a_2 a_3}{\Theta},
\ee
which in general depends on the parameters $a_i$. For simplicity, we confine our search to the case where $G_i = G$, i.e. they are all equal. After changing frame to Einstein frame, consistency of the three equations (\ref{U13gauge2}) then places constraints on $a_i$:
\be
\{a_1 = a_2= a_3 \}, ~~\{a_1= a_2 = \tfrac{2}{7} a_3 \}, ~~\{a_1= a_3 = \tfrac{2}{7} a_2 \}, ~~\{a_2= a_3 = \tfrac{2}{7} a_1 \}.
\ee
Combining these with the condition for a supersymmetric vacuum (\ref{susycond}), one reaches the conclusion that good $AdS_3$ solutions exist only for $\Sigma_{\frak{g}} = H^2$  \footnote{One can compare the values of $a_i$ against Figure 1 of \cite{Benini:2013cda}.}.  The two independent choices we find are
\bea
(a_1, a_2, a_3) &=& (\tfrac{1}{3}, \tfrac{1}{3}, \tfrac{1}{3}), \quad  (a_1, a_2, a_3) = (\tfrac{7}{11}, \tfrac{2}{11}, \tfrac{2}{11}),
\eea
where one is free to consider various cyclic permutations of the latter. The first choice leads to the non-integral value $z = \frac{4}{3}$ with $\ell = \frac{2}{9}$. The second choice does produce an integer, namely $z=18$ with $ \ell = \frac{8}{11}$. Thus, within the limited scope of our search, we do not find any null-warped $AdS_3$ ($z=2$) solutions here.

Moving on, we can turn to the gauged supergravity corresponding to twisted compactifications of $\mathcal{N} =1$ SCFTs, namely (\ref{N1twist}). A particular case of this we have already covered above. Referring the reader to equations (\ref{H2KE4eq1}) and (\ref{H2KE4eq2}), if one truncates consistently to just $K_1, K_2$ and $F_2$, and regardless of how one further truncates to an equation bearing resemblance to (\ref{CSeq}), one finds the dynamical exponent $z= \frac{4}{3}$. This should not come as a surprise as once one truncates to these fields, the theory should correspond to five-dimensional $U(1)^3$ theory where one identifies two of the gauge fields and truncates out a scalar.

However, for the action (\ref{N1twist}), we do have other options. As we are considering a null space-time, it is consistent to truncate to just the scalar $c$ and one-form $C_1$ with the various other scalars taking their vacuum values. Obviously, this is not a consistent truncation in general, but since we assume $G_2 \wedge * G_2 = G_1 \wedge * G_1 = M_1 \wedge * M_1 =0$ in this case, we do not have to worry about the consistency of equations such as (\ref{H2KE4eq5}), (\ref{H2KE4eq3}) and (\ref{H2KE4eq4}). Note that $M_1$ is not independent and is related to $G_1$, $M_1 = \frac{\epsilon}{2} G_1$. This in turn means that, in addition to the Einstein equation, we only have two flux equations
\be
\dd * G_1 =0, \quad \dd * G_2 - \frac{9}{\ell} * G_1 = 0,
\ee
where we have used $\ell = \frac{2}{9}$ and $e^{2C} = \frac{1}{3}$. If we further truncate to set $ * G_1 = -\frac{2}{9} G_2$, then we can find null-warped $AdS_3$ solutions with $z=2$. This allows us to determine $c$ which can be set consistently to zero. In the notation of section \ref{sec:N1twist}, the solution may be expressed as
\bea
\label{nullwarp}
\dd s^2 &=& \ell^2 \left( - r^z \dd u^2 + 2 r \dd u \dd v + \frac{\dd r^2}{4 r^2}\right), \nn
C_1 &=& \tfrac{2}{3} \ell \, r \, \dd u,
\eea
where we have rescaled $C_1$ so that $\lambda = 1$.

We can also consider deformations for $AdS_3$ supported by the scalar $b$ and one-form $B_1$. This involves consistently truncating the action (\ref{N1twist}) to $N_1, H_1$ and $H_2$ and since this may be regarded as the S-dual of the truncation presented immediately above, we recover the same solution.

For some sense of completeness, we also touch upon the existence of solutions for the theory arising from a dimensional reduction on $S^2 \times T^4$ from ten dimensions presented in section \ref{sec:D3}. Schr\"{o}dinger solutions based on the D1-D5 near-horizon, or its S-dual F1-NS5, have already been the focus of considerable attention in the literature. Not only have solutions been constructed directly in ten dimensions \cite{Bobev:2011qx}, but examples in the three-dimensional setting have also been identified in \cite{Detournay:2012dz}. Though not mentioned in \cite{Detournay:2012dz}, an S-duality transformation is all that is required to generate an example supported purely by the NS sector provided one starts with the RR supported two-parameter family of \cite{Detournay:2012dz}. Rather than take this path, we will work directly with our reduced theory and employ an appropriate Ansatz. We will also make use of a further truncation.

Starting from the action in section \ref{sec:D3}, we take $\alpha = \frac{\pi}{2}$ and truncate out various fields $U = V = \phi_i = \chi_i = a = c = C_1 =0$. This corresponds to setting the scalars to their $AdS_3$ vacuum ($\ell =1$) values and the choice of $\alpha$ is appropriate for a vacuum supported solely by NS flux. Further truncating out $A_1$ leads to the condition $ * H_1 = H_2$, leading to the equations of motion:
\bea
\dd * H_2 &=& -2 H_2, \nn
R_{\mu \nu} &=& -2 g_{\mu \nu} + H_{2 \mu \rho} H_{2\nu}^{~~\rho},
\eea
where we have used the fact that $B_1$ is null. Note that the CS equation is now in the accustomed form (\ref{CSeq}), so we can be confident we have a null-warped solution. It is then a straightforward exercise to provide the explicit form of the solution that satisfies these equations of motion:
\bea
\dd s^2 &=& - r^z \dd u^2 + 2 r \dd u \dd v + \frac{\dd r^2}{4 r^2}, \nn
B_1 &=& r \, \dd u.
\eea

It would be interesting to see if any solutions can be generated through applying TsT \cite{Lunin:2005jy} transformations, such as those considered in \cite{Bena:2012wc}.

\section{Outlook}
\label{sec:outlook}
Our primary motivation for this work stems from \cite{Karndumri:2013iqa} where five-dimensional $U(1)^3$ gauged supergravity was dimensionally reduced on a Riemann surface and the lower-dimensional theory re-expressed in terms of the language of three-dimensional gauged supergravity \cite{deWit:2003ja}. As explained in section \ref{sec:generic}, the $T$ tensor presents a natural supergravity counterpart to the quadratic trial function for the central charge presented in \cite{Benini:2012cz, Benini:2013cda} and it is a striking feature that the $T$ tensor, through the embedding tensor, knows about the exact R symmetry. Without recourse to the higher-dimensional solution, this provides a natural way to identify the exact central charge and R symmetry directly in three dimensions.

Since any solution to this particular three-dimensional gauged supergravity uplifts to the $U(1)^3$ theory in five dimensions, which is itself a reduction of type IIB supergravity \cite{Cvetic:1999xp}, we have also taken the opportunity to step back and address consistent KK reductions to three dimensions for wrapped D3-brane geometries. As reviewed in section \ref{sec:wrappedD3}, the origin of supersymmetric $AdS_3$ geometries in type IIB can be traced to D3-branes wrapping K\"ahler two-cycles in Calabi-Yau manifolds, with CFTs of interest to $c$-extremization, namely those with $\mathcal{N} = (0,2)$ supersymmetry, resulting when a two-cycle in a Calabi-Yau four-fold is wrapped. All $AdS_3$ solutions of this form fall into the general classification of supersymmetric geometries presented in \cite{Kim:2005ez} and at the heart of each supersymmetric geometry is a six-dimensional K\"{a}hler manifold $\mathcal{M}_6$, satisfying the differential condition (\ref{diffM6}).

Not only does this condition appear in the flux equations of motion, but the Einstein equation is satisfied through imposing this condition. This makes the task of finding a fully generic KK reduction, in contrast to the case studied in section \ref{sec:generic}, where one assumes the presence of a Riemann surface, an inviting problem. It is expected that one can gauge the $U(1)$ R symmetry and reduce to three dimensions in line with the conjecture of \cite{Gauntlett:2007ma} that gaugings of R symmetry groups always lead to consistent reductions to lower-dimensional gauged supergravities. What is not clear at this moment is whether a truly ``generic" reduction - one working at the level of the supersymmetry conditions - on $\mathcal{M}_6$ exists, thus mimicking general reductions to five dimensions discovered in \cite{Gauntlett:2006ai, Gauntlett:2007ma}, or whether one needs to specify more structure for the $\mathcal{M}_6$. An added subtlety here is that since the reduced theory is expected to fit into $\mathcal{N} =2$ gauged supergravity, it is not enough simply to retain a gauge field coming from an R symmetry gauging and an extra degree of freedom is required.

Naturally enough, what we have discussed here just pertains to D3-branes and $AdS_3$ vacua also arise in eleven-dimensional supergravity arising from wrapped M5-branes. It is then fitting to consider KK reductions from eleven dimensions to three-dimensional gauged supergravity. While supersymmetric $AdS_3$ solutions can be found by considering twists of seven-dimensional supergravity \cite{ Benini:2013cda, Gauntlett:2000ng, Gauntlett:2001jj}, more general solutions are expected to fit into the general classification of supersymmetric solutions presented in \cite{Gauntlett:2006ux, Figueras:2007cn}. A particular case discussed in \cite{Benini:2013cda}, namely seven-dimensional supergravity reduced on $H^2 \times H^2$, we have already considered\footnote{It corresponds to $\mathcal{N} =2$ supergravity with K\"ahler manifold $[SU(1,1)/U(1)]^4$.} and we will report on M5-brane analogues in future work \cite{work}.

In addition to the $c$-extremization angle, another thread to our story concerns the search for null-warped $AdS_3$ or Schr\"{o}dinger ($z=2$) solutions. While it is likely that we have recovered some known solutions, and found solutions with more general $z$,  we believe that the solutions based on $H^2 \times KE_4$ internal geometries are new. What remains is to check whether they preserve supersymmetry, and indeed the identification of the Killing spinor equations for the reduced theories in sections \ref{sec:N1twist} and \ref{sec:D3} needs to be considered  if one is to discuss supersymmetric solutions. The reduction in section \ref{sec:generic} aside, we have simply focused on the bosonic sector and the equations of motion. It may also be interesting to study families of Schr\"odinger solutions interpolating between the D1-D5 vacuum and F1-NS5 vacuum directly in three dimensions. This would presumably overlap with the higher-dimensional examples presented in \cite{Bobev:2011qx}. It is expected that some supersymmetry is preserved.

Combining the principle of $c$-extremization \cite{Benini:2012cz, Benini:2013cda}, which can be understood in terms of three-dimensional supergravity \cite{Karndumri:2013iqa}, and the fact that null-warped $AdS_3$ solutions clearly exist, it is worth considering if $c$-extremization can be extended to warped $AdS_3$. The most immediate setting to address this question is the theory of section \ref{sec:generic}, however, as we have seen, the simplest solutions appear to preclude solutions with $z=2$. A more thorough search for null-warped solutions is warranted. If they do not exist, one can imagine starting from a more involved theory in five dimensions that includes the $U(1)^3$ gauged supergravity. Evidently, the more involved reductions based on $H^2 \times KE_4$ and $S^2 \times T^4$ allow solutions, so it can be expected that this question can be addressed in future work.

It would equally be interesting to look for a holographic analogue of $c$-extremization in two dimensions\footnote{We are grateful to N. Halmagyi for suggesting this possibility.}.  Starting from eleven dimensions, one can reduce to four dimensions \cite{Cvetic:1999xp} retaining the Cartan subgroup $U(1)^4$ of the R symmetry group. Relevant solutions are already  known \cite{Cucu:2003yk, Cucu:2003bm}, and the two-dimensional theory one gets from twisted compactifications on Riemann surfaces are likely to be in the literature, for example \cite{Ortiz:2012ib}, and may be related to BFSS matrix quantum mechanics \cite{Banks:1996vh}. At a quick glance, it looks like we have some of the jigsaw pieces in place.

One of the potentially interesting avenues for future study is to explore the connection between supersymmetric black holes in five dimensions and null-warped $AdS_3$ space-times. For non-relativistic geometries with $z=4$, it was noted in \cite{Colgain:2010rg} that these geometries naturally appear when one considers a general class of five-dimensional supersymmetric black holes and strings and then reduces on an $S^2$. The corresponding picture for the known null-warped solutions can also be worked out. It would be interesting to extend recent studies of the classical motion of strings in warped $AdS_3$ backgrounds \cite{Kameyama:2013qka} to higher-dimensional black holes.

Finally, we are aware of string theory embeddings of holographic superconductors in four and five dimensions \cite{Gubser:2009qm,Gauntlett:2009dn, Gauntlett:2009bh}, where an important element in the construction is the presence of charged scalars that couple to the complex form of the internal K\"{a}hler-Einstein manifold. To date, there is no example of an embedding of the bottom-up model considered in \cite{Ren:2010ha}, though strong similarities between the supersymmetric geometries here and Sasaki-Einstein manifolds suggest that this may be a good place to look. So far we have been unable to find a consistent reduction based on $\mathcal{M}_6 = S^2 \times T^4$ or $\mathcal{M}_6 = H^2 \times KE_4$, but one could hope to address the problem perturbatively. Such an approach was adopted in \cite{Denef:2009tp}.

\section*{Acknowledgements}
We would like to thank P. Szepietowski for sharing his interest in generic reductions at an initial stage of this project. We are grateful to N. Bobev, J. P. Gauntlett, N. Halmagyi, J. Jeong, H. Lu and Y. Nakayama for correspondence on related topics. We have also enjoyed further discussions with M. Guica, E. Sezgin and O. Varela. E \'O C wishes to thank the Centro de Ciencias de Benasque Pedro Pasqual and the organisers of the String Theory meeting for hospitality while we were writing up. P. K. is supported by Chulalongkorn University through the Ratchadapisek Sompote Endowment Fund, Thailand Center of Excellence in Physics through the ThEP/CU/2-RE3/11 project, and The Thailand Research Fund (TRF) under grant TRG5680010. E. \'O C acknowledges support from the research grant MICINN-09- FPA2012-35043-C02-02.

\appendix

\section{Type IIB supergravity conventions}
\label{sec:IIBsugra}
Our conventions for type IIB supergravity follow those of \cite{Gauntlett:2010vu}, which for completeness, we reproduce here. Restricting ourselves to the bosonic sector of type IIB supergravity, the field content consists of RR $n$-forms $F_{(n)}$, $n=1, 3, 5$, the NS form $H_{(3)}$, the dilaton $\Phi$ and the metric. The forms satisfy the Bianchi identities
\bea
&& \dd F_{(5)} + F_{(3)} \wedge H_{(3)} = 0, \\
&& \dd F_{(3)} + F_{(1)} \wedge H_{(3)} = 0, \\
&& \dd F_{(1)} = 0, \\
&& \dd H_{(3)} = 0,
\eea
which can be satisfied through the introduction of potentials $C_{(n-1)}$, $B_{(2)}$. In terms of these potentials, the forms are $F_{(5)} = \dd C_{(4)} - C_{(2)} \wedge H_{(3)}, F_{(3)}= \dd C_{(2)} - C_{(0)} H_{(3)}, F_{(1)} = \dd C_{(0)}$, $H_{(3)} = \dd B_{(2)}$. In addition to the self-duality condition on the five-form, $* F_{(5)} = F_{(5)}$, the equations of motion take the form:
\bea
&& \dd ( e^{\Phi} * F_{(3)}) - F_{(5)} \wedge H_{(3)} = 0, \\
&& \dd (e^{2 \Phi} * F_{(1)}) + e^{\Phi} H_{(3)} \wedge * F_{(3)} = 0, \\
&& \dd (e^{-\Phi} * H_{(3)} ) - e^{\Phi} F_{(1)} \wedge * F_{(3)} - F_{(3)} \wedge F_{(5)} = 0, \\
&& \dd * \dd \Phi - e^{2 \Phi} F_{(1)} \wedge * F_{(1)} + \tfrac{1}{2} e^{-\Phi} H_{(3)} \wedge * H_{(3)} - \tfrac{1}{2} e^{\Phi} F_{(3)} \wedge * F_{(3)} = 0, \\
&& R_{MN} = \tfrac{1}{2} \partial_{M} C_{(0)} \partial_{N} C_{(0)} + \tfrac{1}{2} \partial_{M} \Phi \partial_N \Phi + \tfrac{1}{96} F_{M PQRS} F_{N}^{~PQRS} \nn
&& \phantom{xxxxxx} \tfrac{1}{4} e^{-\Phi} (H_{M}^{~PQ} H_{NPQ} - \tfrac{1}{12} g_{MN} H^{PQR} H_{PQR} ), \nn
&& \phantom{xxxxxx} \tfrac{1}{4} e^{\Phi} (F_{M}^{~PQ} F_{NPQ} - \tfrac{1}{12} g_{MN} F^{PQR} F_{PQR} ).
\eea

\section{Connection between \cite{Colgain:2010rg} and \cite{Detournay:2012dz}}
\label{sec:3dreds}
In this section we will discuss the connection between two dimensional reductions from higher-dimensional supergravities to three-dimensional theories that have appeared in the literature. Both theories admit supersymmetric Schr\"{o}dinger solutions, however, for those based on the D1-D5 near-horizon \cite{Detournay:2012dz} the dynamical exponent $z=2$ appears, while the dynamical exponent quoted in \cite{Colgain:2010rg} is $z=4$.

Recall that these theories support $AdS_3$ vacua whose higher-dimensional manifestations are $AdS_3 \times S^3 \times CY_2$ geometries of type IIB supergravity and $AdS_3 \times S^2 \times CY_3$ geometries of eleven-dimensional supergravity, respectively. Specialising to the case where the Calabi-Yau  three-fold is a direct product involving a torus $T^2$, $CY_3 = CY_2 \times T^2$,  it is a well-known fact that the geometries are related via dimensional reduction and T-duality. This raises a question about the difference in the quoted dynamical exponents. Here we address that issue and show that a sub-truncation of \cite{Detournay:2012dz} and \cite{Colgain:2010rg} is common and that amongst the $z=2$ solutions presented in \cite{Detournay:2012dz}, one can also find a $z=4$ solution.

We start by considering the KK reduction Ansatz from eleven-dimensions. The solution appearing in \cite{Colgain:2010rg} has a higher-dimensional manifestation of the form
\bea
\dd s^2_{11} &=& e^{-4 W} \dd s^2_3+ e^{2W} \dd s^2(S^2) + \dd s^2(CY_2) + \dd x_5^2 + \dd x_6^2, \nn
G_4 &=& (\alpha \vol(S^2) + H_2) \wedge \left( J_{CY_2} + \dd x_5 \wedge \dd x_6 \right),
\eea
where we have consistently truncated out the fields $f, V, B_1$ leaving just a scalar $W$ and one-form potential $B_2$, where $H_2 = \dd B_2$. Here $(x_5, x_6)$ label coordinates on the $T^2$ and $\alpha$ is a constant. Plugging this Ansatz into the equations of motion of eleven-dimensional supergravity one finds \cite{Colgain:2010rg}
\bea
\label{eom1} \dd (e^{4 W} *_3 H_2 ) &=& - 2 \alpha H_2,  \\
\label{eom2} \dd *_3  \dd W &=& \tfrac{1}{2} e^W H_2 \wedge *_3 H_2 + (e^{-6 W} - \alpha^2 e^{-8 W} ) \vol_3,
\eea
and the Einstein equation which we omit.

Dimensional reduction on $x_6$ and T-duality on $x_5$ leads to the following IIB KK reduction Ansatz
\bea
\dd s^2_{10} &=& e^{-4 W} \dd s^2_3 + e^{2W} \dd s^2(S^2) + \dd s^2(CY_2) +  (\dd x_5 - \alpha \cos \theta \dd  \phi + B_2)^2, \\
F_5 &=& (1+*_{10}) \biggl[\alpha \vol(S^2) \wedge J_{CY_2} + J_{CY_2} \wedge H_2 \biggr] \wedge (\dd x_5 - \alpha \cos \theta \dd \phi + B_2), \nonumber
\eea
where $(\theta, \phi)$ parametrise the two-sphere $S^2$ and all other fields, including the dilaton are zero.

At this point it is easier to compare with the ten-dimensional uplift \cite{Duff:1998cr} of the six-dimensional Ansatz considered in \cite{Detournay:2012dz} to get our bearings. After rescaling the metric to make the transition to string frame, the ten-dimensional space-time may be written as
\bea
\dd s^2_{10} &=& e^{\frac{\phi_1}{2} + \frac{\phi_2}{2} } \dd s^2_6 + e^{\frac{\phi_1}{2} - \frac{\phi_2}{2}} \dd s^2(CY_2), \nn
\dd s^2_6 &=& e^{-4 U -2 V} \dd s^2_3 + \tfrac{1}{4} e^{2 U} \dd s^2(S^2) + \tfrac{1}{4} e^{2V} ( \dd \psi + \cos \theta \dd \phi + 2 A)^2,
\eea
where we have set the length-scale $\ell$ corresponding to the $AdS_3$ radius to unity for simplicity. To compare the metrics we note that we require the following identifications:
\bea
\phi = \phi_1 = \phi_2 = -2 V, \quad e^{W} = \tfrac{1}{2} e^{U}, \quad 2 x_5 = \psi, \quad \alpha = - \tfrac{1}{2}, \quad B_2 = A.
\eea
While this places us in the class of consistent reductions in section 4.2 of \cite{Detournay:2012dz}, the added condition that the dilaton $\phi$ is zero tells us that the scalars $\phi, V$ appearing in equations (B.25) and (B.29) of \cite{Detournay:2012dz} are zero. These equations together then tell us that the two gauge fields appearing in \cite{Detournay:2012dz} should be identified $A = \pm \hat{A}$. For $CY_2 = T^4$, the RR-sector is then simply related via T-duality.

The choice $A = \hat{A}$ immediately leads to the condition $F^2 = 0$ through (B.25), however there is another option. We can choose $A= - \hat{A}$ with the further relation
\be
A = \tfrac{1}{4} e^{4 U} *_3 F.
\ee
With this relation one can then satisfy oneself that (B.27) and the $U$ equation from (B.29) of \cite{Detournay:2012dz} can be identified with (\ref{eom1}) and (\ref{eom2}) above, meaning that this particular sub-truncation of both reductions is the same.

Indeed, since the higher-dimensional $AdS_3$ solutions can be related via dimensional reduction and T-duality, it is expected that the KK reductions are also related at some level.


\section{Details of reduction of $D=5$ $U(1)^3$ gauged supergravity}
\label{sec:U13}

Here we begin by recording the five-dimensional equations of motion one gets from varying the action (\ref{U13act}). The equations of motion for the gauge fields $A_i$, $i=1, 2, 3,$
are
\bea \dd (X_1^{-2}
* F^1) &=& F^2 \wedge F^3, \nn \dd (X_2^{-2} * F^2) &=& F^1 \wedge
F^3, \nn \dd (X_3^{-2} * F^3) &=& F^1 \wedge F^2,
\eea
and those of the scalars are given by
\bea \dd * \dd \varphi_1 &=& \tfrac{1}{\sqrt{6}}
\left( X_1^{-2} F^1 \wedge * F^1 +  X_2^{-2} F^2 \wedge * F^2 - 2
X_3^{-2} F^3 \wedge * F^3   \right)  \nn
&-& g^2 \tfrac{4}{\sqrt{6}} \left( X_1^{-1} + X_2^{-1} - 2 X_3^{-1}\right) \vol_5, \\
\dd * \dd \varphi_2 &=& \tfrac{1}{\sqrt{2}} \left( X_1^{-2} F^1 \wedge * F^1 -  X_2^{-2} F^2 \wedge * F^2  \right) - g^2 2 \sqrt{2} \left( X_1^{-1} - X_2^{-1} \right) \vol_5. \nonumber
\eea
\\
\indent Finally, the Einstein equation reads
\bea R_{\mu \nu} &=&
\tfrac{1}{2} \sum_{i=1}^2 \partial_{\mu} \varphi_i \partial_{\nu}
\varphi_i + \tfrac{1}{2} \sum_{i=1}^3 X_i^{-2} \left( F^i_{\mu \rho}
F^{i~\rho}_{\nu} - \tfrac{1}{6} g_{\mu \nu} F^i_{\rho \sigma} F^{i \,
\rho \sigma} \right) \nn &-& g_{\mu \nu} \tfrac{4}{3} g^2 \sum_{i=1}^3
X_i^{-1}. \eea

The reduction at the level of the equations of motion is most simply performed be first reducing on the internal space, in this case a Riemann surface $\Sigma_{\frak{g}}$, and then rescaling the external space-time to go to Einstein frame. Thus, here we consider the initial Ansatz for the five-dimensional space-time
\be \dd s^2_5 = \dd s^2_3 + e^{2C} \dd s^2 (\Sigma_{\frak{g}}), \ee where $C$ is a
scalar warp factor depending on the coordinates of the
three-dimensional space-time.

To reduce the gauge field strengths we consider the Ansatz (\ref{U13gauge}). The equations of motion for the gauge fields now reduce as
\bea
\label{U13gauge2}
\dd \left( X_1^{-2} e^{2 C}*_3 G^1 \right)  &=& -(a_3 G^2 + a_2 G^3 ),
\nn \dd \left( X_2^{-2} e^{2 C} *_3 G^2 \right)  &=& -(a_3 G^1 + a_1
G^3 ), \nn \dd \left( X_3^{-2} e^{2 C} *_3 G^3 \right)  &=& -(a_1 G^2
+ a_2 G^1 ). \eea

From the scalar equations of motion, we find \bea \dd ( e^{2C}
*_3 \dd \varphi_1) &=& \tfrac{1}{\sqrt{6}} e^{2C} \biggl[   X_1^{-2}
\left( G^1 \wedge *_3 G^1 + a_1^2 e^{-4C} \vol_3 \right)  + X_2^{-2}
\biggl(  G^2 \wedge
*_3 G^2 \nn &+& a_2^2 e^{-4 C} \vol_3 \biggr) - 2 X_3^{-2} \biggl(
G^3 \wedge *_3 G^3 + a_3^2 e^{-4 C} \vol_3 \biggr) \biggr]\nn &-&
g^2 \tfrac{4}{\sqrt{6}} e^{2C} \left( X_1^{-1} + X_2^{-1} - 2
X_3^{-1}  \right) \vol_3,  \nn
\dd ( e^{2C} *_3 \dd \varphi_2) &=&
\tfrac{1}{\sqrt{2}} e^{2 C} \biggl[  X_1^{-2} \left( G^1 \wedge *_3
G^1 + a_1^2 e^{-4C} \vol_3 \right)  - X_2^{-2} \biggl(  G^2 \wedge
*_3 G^2 \nn &+& a_2^2 e^{-4 C} \vol_3 \biggr)\biggr] - 2 \sqrt{2}
g^2 e^{2 C} \left(X_1^{-1} -  X_2^{-1}\right) \vol_3. \eea

The Einstein equation along the Riemann surface presents us with
another scalar equation of motion, this time for $C$: \bea -
\nabla_{\mu} \nabla^{\mu} C - 2 \partial_{\mu} A \partial^{\mu} C +
e^{-2C} \kappa &=& \tfrac{1}{2} \sum_{i=1}^3 X_i^{-2} \left( \tfrac{2}{3}
a_i^2 e^{-4C} - \tfrac{1}{6} G^i_{\rho \sigma} G^{i \, \rho \sigma}
\right)  \nn &-& \tfrac{4}{3} g^2 \sum_{i=1}^3 X_i^{-1}, \eea
where $\kappa$ is the curvature of the Riemann surface.

Finally, the Einstein equation in three dimensions may be written as
\bea R_{\mu \nu} &=& 2 ( \nabla_{\nu} \nabla_{\mu} C +
\partial_{\mu} C \partial_{\nu} C ) + \sum_{i=1}^2 \partial_{\mu}
\varphi_i \partial_{\nu} \varphi_i + \tfrac{1}{2} \sum_{i=1}^3 X_i^{-2}
\left( G^i_{\mu \rho} G^{i~\rho}_{\nu} - \tfrac{1}{6} g_{\mu \nu}
G^i_{\rho \sigma} G^{i \, \rho \sigma} \right) \nn &-& \tfrac{1}{6}
g_{\mu \nu} \sum_{i=1}^3 \left( a_i^2 e^{-4C} X_i^{-2} + 8 g^2 X_i^{-1}
\right). \eea

The above equations can be shown to result from varying the action
\bea \label{act1}
\mathcal{L}^{(3)} &=& e^{2C} \biggl[ R *_3 \mathbf{1} + 2 \dd C \wedge *_3
\dd C - \tfrac{1}{2} \sum_{i=1}^2 \dd \varphi_i \wedge *_3 \dd \varphi_i  -
\tfrac{1}{2} \sum_{i=1}^3 X_i^{-2} G^i \wedge *_3 G^i\biggr]\nn &+&
\left( \sum_{i=1}^3 \left[ 4 g^2 e^{2C} X_i^{-1} - \tfrac{1}{2} e^{-2C}
a_i^2 X_i^{-2} \right]  + 2 \kappa \right) *_3 \mathbf{1} +
\mathcal{L}^{(3)}_{\textrm{top}}, \eea
where the topological term is \be
\mathcal{L}^{(3)}_{\textrm{top}} = a_1 B^2 \wedge G^3 + a_2 B^3 \wedge G^1
+ a_3 B^1 \wedge G^2\, . \ee
Here $B^i$ is the one-form potential for $G^i$,
$G^i = \dd B^i$.

Now, to go to Einstein frame we just need to do a
conformal transformation, $g_{\mu \nu} = e^{-4 C} \hat{g}_{\mu \nu}$. This leads to the Einstein frame action  (\ref{Einsteinact}) quoted in the text.

In checking the Einstein equation we have made use of the following Ricci tensor components
\bea
R_{\mu \nu} &=& \bar{R}_{\mu \nu} - 2 ( \nabla_{\nu} \nabla_{\mu} C + \partial_{\nu} C \partial_{\mu} C ), \nn
R_{mn} &=& \left[ \kappa e^{-2C} - \nabla_{\mu} \nabla^{\mu} C - 2 \partial_{\mu} C \partial^{\mu} C \right] \delta_{mn},
\eea
where $\mu, \nu$ label space-time directions and $m, n$ correspond to directions on the Riemann surface.

\subsection{Killing spinor equations}
\label{sec:KSE}
We would like to confirm that the $T$ tensor (\ref{T}) can be extracted directly from the Killing spinor equations via reduction. In a related context, a similar calculation appeared
in \cite{Szepietowski:2012tb} and in that context assisted the identification of a five-dimensional prepotential. Our motivation here is the same.

We adopt the conventions for the Killing spinor equations in $D=5$ from (F.1) of
\cite{Benini:2013cda} (see also \cite{Maldacena:2000mw}), and in some sense, up to some additional fields, the calculation here is almost identical to appendix F of \cite{Benini:2013cda}. We work with the natural vielbein
 \be
e^{\mu} = e^{-2C} \bar{e}^{\mu}, \quad e^{a} = e^{C} \bar{e}^a,
\ee
where $\mu = 0, 1, 2$ label three-dimensional space-time directions and $a = 3, 4$ denote directions
along the Riemann surface. Our Ansatz for the flux follows from (\ref{U13gauge}).

For the Killing spinor we make the choice
 \be \e = e^{\beta C} \xi \otimes \eta, \ee
 where $\beta$ is a constant we will fix later. We use the following
decomposition of the five-dimensional gamma matrices \be \gamma^{\mu} = \rho^{\mu}
\otimes \s^3, \quad \gamma^3 = 1 \otimes \s^1, \quad \gamma^4 = 1
\otimes \s^2. \ee As in \cite{Benini:2013cda}, where one has
$\gamma_{34} \e = i \e$, following decomposition, we have $\s^3 \eta = \eta$.

Inserting the Ansatz into the Killing spinor equations we arrive at
\bea \label{phi0} 2 \delta
\psi_{3} &=& \left[\g_{3}^{~\mu} e^{2C} \partial_{\mu} C +
\sum_{i=1}^3 \biggl(X_i \g_3 + \tfrac{i}{3} e^{-2C}
a_i X_i^{-1} \g_4 + \tfrac{i}{12}
e^{4C} \g_{3}^{~\mu \nu} X_i^{-1} G^i_{\mu
\nu} \biggr) \right]e^{\beta C}\xi\otimes \eta,
\nn
\\
\label{phi1} \sqrt{6} \delta \chi_{(1)} &=& \biggl[
\tfrac{1}{8}\sum_{i=1}^2 X_i^{-1} \left( e^{4C} G^i_{\mu \nu} \g^{\mu
\nu} - 2i a_i e^{-2C} \right) - \tfrac{1}{4} X_3^{-1} \left( e^{4C}
G^3_{\mu \nu}
\g^{\mu \nu}  - 2i a_3 e^{-2C} \right) \nn& & +\tfrac{i}{2} \left(-X_1 - X_2 + 2 X_3\right) - i \tfrac{\sqrt{6}}{4} e^{2C} \partial_{\mu} \varphi_1 \gamma^{\mu} \biggr] e^{\b C} \xi \otimes \eta, \\
\label{phi2} \sqrt{2} \delta \chi_{(2)} &=& \biggl[
\tfrac{1}{8} X_1^{-1} \left(e^{4C} G^1_{\mu \nu} \g^{\mu
\nu} - 2i a_1 e^{-2C} \right) -\tfrac{1}{8} X_2^{-1} \left(e^{4C} G^2_{\mu \nu} \g^{\mu
\nu} - 2i a_2 e^{-2C} \right) \nn && +\tfrac{i}{2} \left(-X_1 + X_2
\right) - i \tfrac{\sqrt{2}}{4} e^{2C}
\partial_{\mu} \varphi_2 \gamma^{\mu} \biggr] e^{\b C} \xi \otimes
\eta. \eea

Note, in contrast to \cite{Benini:2013cda} where scalars with raised and lowered
indices are employed,  here our $X_i$ are simply those in (\ref{x}).
As a consistency check, (\ref{phi0}), (\ref{phi1}),
(\ref{phi2}) agree with (3.20) of \cite{Benini:2013cda} when $G^i
=0$ and $\phi_i = \phi_i(r), C = g(r)$.

Taking various linear combinations we can write
\bea
 4 \g^3 \delta \psi_3 + \tfrac{2}{3} \sqrt{6} i \delta \chi_{(1)} + 2 \sqrt{2} i \delta \chi_{(2)}&=& \delta_{\epsilon} \lambda^1 \otimes  \eta, \nn
 4 \g^3 \delta \psi_3 + \tfrac{2}{3} \sqrt{6} i \delta \chi_{(1)} - 2 \sqrt{2} i \delta \chi_{(2)}&=& \delta_{\epsilon} \lambda^2 \otimes  \eta, \nn
 4 \g^3 \delta \psi_3  - \tfrac{4}{3} \sqrt{6} i \delta \chi_{(1)} &=& \delta_{\epsilon} \lambda^3 \otimes  \eta
\eea
leading to the variations (constant $\beta = -2$)
\bea
\delta_{\epsilon } \lambda^1 &=& \left[ \rho^{\mu} \partial_{\mu} W_1 + \tfrac{i}{2} X_1^{-1} e^{2C}  G^1_{\mu \nu} \rho^{\mu \nu}  + e^{-4 C} \left( 2 e^{2C} X_1 - a_2 X_2^{-1} -a_3 X_3^{-1} \right) \right]
\xi, \nn
\delta_{\epsilon } \lambda^2 &=& \left[ \rho^{\mu} \partial_{\mu} W_2 + \tfrac{i}{2} X_2^{-1} e^{2C}  G^2_{\mu \nu} \rho^{\mu \nu}  + e^{-4 C} \left( 2 e^{2C} X_2 - a_1 X_1^{-1} -a_3 X_3^{-1} \right) \right]
\xi, \nn
\delta_{\epsilon } \lambda^3 &=& \left[ \rho^{\mu} \partial_{\mu} W_3 + \tfrac{i}{2} X_1^{-3} e^{2C}  G^3_{\mu \nu} \rho^{\mu \nu}  + e^{-4 C} \left( 2 e^{2C} X_3 - a_1 X_1^{-1} -a_2 X_2^{-1} \right) \right]
\xi.
\eea
Dualising $G^i$ as instructed in the text, the above equations can be condensed into a single equation
\be
\delta_{\epsilon } \lambda^a = 2 E_{i}^{~a} \left( \rho^{\mu}  D_{\mu} z^i -  2 \partial^{i} T \right),
\ee
which is the expected form for the Killing spinor equation for the spinor fields \cite{deWit:2003ja,Colgain:2010rg} and we see that the $T$ tensor (\ref{T}) features. $E_{i}^{~a}$, $a=1, 2, 3,$ is the complex dreibein defined through $g_{i \bar{i}} = E_{i}^{~a} E_{\bar{i} a}$, where $E_{\bar{i} a} = (E_{i}^{~a})^*$.

\section{Curvature for K\"{a}hler-Einstein space-times}
\label{sec:KE6ricci}
Working in Einstein frame, we adopt the following Ansatz for the space-time
\be
\dd s^2_{10} = e^{2A} \dd s^2 (\mathcal{M}_3)+ e^{2A} \tfrac{1}{4} e^{2 W} (\dd z + P + A_1)^2 + e^{-2A}  \sum_{a=1}^3 e^{2 V_a} \dd s^2(KE_2^{(a)}).
\ee
where $A$ is a \textit{constant} overall factor, we have dropped the overall scale $L$ appearing in (\ref{gensol}) and $W$, $V_a$, $a=1, 2, 3$ denote scalar warp factors. $A_1$ is a one-form living on the three-dimensional space-time $\mathcal{M}_3$.

We adopt the natural orthonormal frame
\be
e^{\mu} = e^{A} \, \bar{e}^{\mu}, \quad e^{z} = e^{A +W} \, \tfrac{1}{2} (\dd z + P + A_1), \quad e^i = e^{-A+V_a} \, \bar{e}^i,
\ee
where $\mu = 0, 1, 2$ label $AdS_3$ directions and $i = 3, \dots, 8$  correspond to directions along the internal K\"{a}hler-Einstein spaces.

With constant $A$, the spin-connection for the metric may be written as
\bea
\label{spinAconst}
\omega^{\mu}_{~\nu} &=& \bar{\omega}^{\mu}_{~\nu} - \tfrac{1}{4} e^{-A+W} (F_2)^{\mu}_{~\nu} e^{z}, \nn
\omega^{i}_{~j} &=& \bar{\omega}^{i}_{~j} - \tfrac{1}{4} e^{3 A + W -2 V_a} l_a (J_a)^i_{~j} e^{z}, \nn
\omega^{\mu}_{~z} &=& - e^{-A} \partial^{\mu} W e^z - \tfrac{1}{4} e^{-A+W} (F_2)^{\mu}_{~\rho} e^{\rho}, \nn
\omega^{i}_{~\mu} &=& e^{-A} \partial_{\mu} V_a e^{i}, \nn
\omega^{i}_{~z} &=& - \tfrac{1}{4} e^{3 A+W - 2 V_a} l_a (J_a)^i_{~j} e^{j}.
\eea

Using the above spin-connection one can calculate the Ricci-form
\bea
\label{RicciAconst}
R_{\mu \nu} &=& e^{-2A} \biggl[ \bar{R}_{\mu \nu} - (\nabla_{\nu} \nabla_{\mu} W + \partial_{\mu} W \partial_{\nu} W) - \sum_{a=1}^3 2 (\nabla_{\nu} \nabla_{\mu} V_a + \partial_{\mu} V_a \partial_{\nu} V_a) \nn && \phantom{xxxxxxxxxxxxxxx} -  \tfrac{1}{8} e^{2 W} F_{2\,\mu \rho} F_{2\,\nu}^{~~\rho} \biggr] , \nn
R_{zz} &=& \tfrac{1}{8} e^{6A +2 W} \sum_{a=1}^3 e^{-4 V_a} l_a^2 - e^{-2A} ( \nabla_{\mu} \nabla^{\mu} W + \partial_{\mu} W \partial^{\mu} W ) - 2 \partial^{\mu} W \sum_{a=1}^3 e^{-2 A} \partial_{\mu} V_a  \nn
&& \phantom{xxxxxxxxxxxxx} + \tfrac{1}{16} e^{-2 A + 2 W} F_{2\,\rho \sigma} F_2^{\rho \sigma}, \nn
R_{11} &=& R_{22} =  e^{-2A} \biggl[ - \nabla_{\mu} \nabla^{\mu} V_1 -  \partial_{\mu} W \partial^{\mu} V_1 - 2 \partial_{\mu} V_1 \sum_{i=a}^3 \partial^{\mu} V_a \biggr] + l_1 e^{2A-2 V_1} - \tfrac{1}{8} l_1^2 e^{4 A + 2 W - 4 V_1}, \nn
R_{33} &=& R_{44} =  e^{-2A} \biggl[ - \nabla_{\mu} \nabla^{\mu} V_2 -  \partial_{\mu} W \partial^{\mu} V_2 - 2 \partial_{\mu} V_2 \sum_{i=a}^3 \partial^{\mu} V_a \biggr] + l_2 e^{2A-2 V_2} - \tfrac{1}{8} l_2^2 e^{4 A + 2 W - 4 V_2}, \nn
R_{55} &=& R_{66} =  e^{-2A} \biggl[ - \nabla_{\mu} \nabla^{\mu} V_3 -  \partial_{\mu} W \partial^{\mu} V_3 - 2 \partial_{\mu} V_3 \sum_{a=1}^3 \partial^{\mu} V_a \biggr] + l_3 e^{2A-2 V_3} - \tfrac{1}{8} l_3^2 e^{4 A + 2 W - 4 V_3}, \nn
R_{\mu z} &=& - \tfrac{1}{4} e^{-2 W - 2 (V_1 + V_2 +V_3)} \nabla_{\rho} ( e^{3 W + 2(V_1 + V_2 + V_3)} F_{2~\mu}^{~\rho} ),
\eea
where all other terms are zero.

\section{Details of reduction on $H^2 \times KE_4$}
\label{sec:H2KE4}
In this section we record equations of motion of the dimensionally reduced three-dimensional theory. This will be useful for testing the consistency of the reduction. We begin with the Bianchi identities. The Bianchi identities for the three-form fluxes $F_{(3)}$ and $H_{(3)}$ are trivially satisfied using the expressions in the text.  The Bianchi for $F_{(5)}$ is partially satisfied, with the remaining equations being:
\bea
\label{H2KE4eq1}
&& \dd (e^{-\frac{4}{3} (U+V) +4 C} * K_2) - 4 e^{-8 U} * K_1 + \epsilon (K_2 - F_2) -N_1 \wedge G_1 - H_1 \wedge M_1 = 0, \nn
&& \dd (e^{-8 U} * K_1) + \tfrac{1}{2} N_1 \wedge G_2 - \tfrac{1}{2} H_2  \wedge M_1 = 0.
\eea
The equations of motion for $F_{(3)}$ and $H_{(3)}$ give respectively the equations
\bea
&& \dd ( e^{\frac{4}{3} (4 U + V) + \phi - 4C} * M_1) -4 h \vol_3 + 2 H_2 \wedge K_1 - 2 H_1 \wedge K_2 = 0, \nn
&& \dd ( e^{\frac{4}{3} (2 U - V) + \phi +4 C} * G_2) - 4 e^{-4 U + \phi} * G_1 - \epsilon e^{\frac{4}{3} (4 U + V) + \phi - 4C} * M_1 \nn &&
\phantom{xxxxxxxxx} + g e^{\frac{4}{3} (4 U + V) + \phi + 8 C} F_2 +2 N_1 \wedge K_1 + 2 e^{- \frac{4}{3} (U + V) + 4C} H_1 \wedge * K_2 = 0, \nn
&& \dd (e^{-4 U + \phi} * G_1 ) - N_1 \wedge K_2 + \epsilon h \vol_3 + e^{-\frac{4}{3} (U+V) + 4C} H_2 \wedge * K_2 \nn && \phantom{xxxxxxxxxxxxxxxxxxxxxxxxxxxxxx}+ 2 e^{-8 U} H_1 \wedge * K_1 = 0,
\eea
and
\bea
&& \dd ( e^{\frac{4}{3} (4 U + V) - \phi -4C} * N_1) + 4 g \vol_3 -2 G_2 \wedge K_1 + 2 G_1 \wedge K_2 - e^{\frac{4}{3} (4 U + V) + \phi - 4 C} \dd a \wedge * M_1 = 0, \nn
&& \dd ( e^{\frac{4}{3} (2 U - V) - \phi +4C} * H_2) - 4 e^{-4 U - \phi} * H_1 - \epsilon e^{\frac{4}{3} (4 U + V) - \phi-4C} * N_1 +  h e^{\frac{4}{3} (4 U + V) - \phi + 8 C} F_2 \nn
&& \phantom{xxxxxxx} - 2 M_1 \wedge K_1 - 2 e^{- \frac{4}{3} (U + V) + 4 C} G_1 \wedge * K_2 -  e^{\frac{4}{3} (2 U - V) + \phi+4C} \dd a \wedge * G_2 = 0, \nn
&& \dd ( e^{-4 U - \phi} * H_1 )  + M_1 \wedge K_2 - \epsilon g \vol_3 - e^{-\frac{4}{3} (U + V) + 4C} G_2 \wedge * K_2 - 2 e^{-8U} G_1 \wedge * K_1 \nn && \phantom{xxxxxxxxxxxxxxxxxxx} - e^{-4 U + \phi} \dd a \wedge * G_1 = 0.
\eea
The axion and dilaton equation are respectively
\bea
&& \dd ( e^{2 \phi} * \dd a) + e^{\frac{4}{3} (4 U + V) + \phi -4 C} N_1 \wedge * M_1 - e^{\frac{4}{3} (4 U +V) + \phi +8 C} g h \vol_3 + e^{\frac{4}{3} (2 U - V) + \phi + 4 C} H_2 \wedge * G_2 \nn && 2 e^{-4 U + \phi} H_1 \wedge * G_1 = 0,
\eea
and
\bea
\label{H2KE4eq5}
&& \dd  * \dd \phi - e^{2 \phi} \dd a \wedge * \dd a + \tfrac{1}{2} e^{\frac{4}{3} (4 U + V) -4 C} \left[ e^{-\phi} N_1 \wedge * N_1 - e^{\phi} M_1 \wedge * M_1 \right] \nn
&& - \tfrac{1}{2} e^{\frac{4}{3} (4 U + V) + 8 C} \left[ e^{-\phi} h^2 - e^{\phi} g^2 \right] \vol_3 + \tfrac{1}{2} e^{\frac{4}{3} (2 U - V) +4 C} \left[ e^{-\phi} H_2 \wedge * H_2 - e^{\phi} G_2 \wedge * G_2  \right] \nn
&& + e^{-4U} \left[ e^{-\phi} H_1 \wedge * H_1 - e^{\phi} G_1 \wedge * G_1 \right] = 0.
\eea
The equations of motion for $A_1, U$ and $V$ are
\bea
\label{H2KE4eq2}
&& \dd (e^{\frac{8}{3} (U+V) +4 C} * F_2) -2 \epsilon K_2 - 8 e^{-8 U} * K_1 + e^{\frac{4}{3} (4 U +V) +8 C} \left[ e^{-\phi} h H_2 + e^{\phi} g G_2 \right] = 0, \phantom{xx}
\eea
\bea
\label{H2KE4eq3}
&& \dd * \dd  U + e^{-8 U} K_1 \wedge * K_1 - \tfrac{1}{8} e^{\frac{4}{3} (4 U + V) -4 C} \left[ e^{- \phi} N_1 \wedge * N_1+ e^{\phi} M_1 \wedge * M_1 \right] \nn
&& + \tfrac{1}{8} e^{\frac{4}{3} (4 U + V) +8 C} \left[ e^{-\phi} h^2 + e^{\phi} g^2 \right] \vol_3 - \tfrac{1}{8} e^{\frac{4}{3} (2 U -V) +4 C} \left[ e^{-\phi} H_2 \wedge * H_2 + e^{\phi} G_2 \wedge * G_2 \right]  \\
&& +\tfrac{1}{4} e^{-4 U} \left[ e^{-\phi} H_1 \wedge * H_1 + e^{\phi} G_1 \wedge * G_1 \right] + e^{-4C} (-6 e^{-\frac{2}{3} (7 U + V)} + 2 e^{\frac{4}{3} (-5 U + V)} + 4 e^{- \frac{8}{3} (4 U + V) } ) = 0, \nonumber
\eea
\bea
\label{H2KE4eq4}
&& \dd  * \dd V  -\tfrac{1}{8} e^{\frac{4}{3} (4 U + V) -4 C} \left[ e^{- \phi} N_1 \wedge * N_1+ e^{\phi} M_1 \wedge * M_1 \right]  + \tfrac{1}{8} e^{\frac{4}{3} (4 U + V) +8 C} \left[ e^{-\phi} h^2 + e^{\phi} g^2 \right] \vol_3 \nn
&& - \tfrac{1}{2} e^{\frac{8}{3} (U + V)+4 C} F_2 \wedge * F_2 + \tfrac{1}{2} e^{- \frac{4}{3} (U + V) +4 C} K_2 \wedge * K_2 - e^{-8 U} K_1 \wedge * K_1 \nn
&& - \tfrac{1}{2} \epsilon^2  e^{\frac{8}{3} (U + V)-8 C} \vol_3 + \tfrac{1}{2} \epsilon^2 e^{- \frac{4}{3} (U + V) -8 C} \vol_3 + \tfrac{3}{8} e^{\frac{4}{3} (2 U -V) +4 C} \left[ e^{-\phi} H_2 \wedge * H_2 + e^{\phi} G_2 \wedge * G_2 \right] \nn
&& - \tfrac{1}{4} e^{-4 U} \left[ e^{-\phi} H_1 \wedge * H_1 + e^{\phi} G_1 \wedge * G_1 \right] + e^{-4C} (-4 e^{\frac{4}{3} (-5 U + V)} + 4 e^{-\frac{8}{3} (4 U + V)})\vol_3 = 0.
\eea

\section{Details of reduction on $S^2 \times T^4$}
\label{sec:CYred}

\subsection*{IIB reduced on $CY_2$}
Here we briefly review the KK reduction Ansatz of type IIB on a Calabi-Yau two-fold that featured in \cite{Duff:1998cr}. The KK Ansatz in Einstein frame is
\bea
\dd s^2_{10} &=& e^{\frac{1}{2} \phi_2 } \dd s^2_6 + e^{- \frac{1}{2} \phi_2} \dd s^2(CY_2), \nn
F_{(5)} &=&  \vol(CY_2) \wedge \dd \chi_2 + e^{2 \phi_2} *_6 \dd \chi_2,
\eea
and all other fields of type IIB supergravity simply reduce to six dimensions. This Ansatz thus leads to extra scalars in addition to the axion $\chi_1$ and dilaton $\phi_1$ of type IIB supergravity, one corresponding to a breathing mode $\phi_2$, and another axion $\chi_2$ coming from the self-dual five-form flux. The six-dimensional action is
\bea
\label{action6d} e^{-1} \mathcal{L} &=& R - \sum_{i=1}^2 \tfrac{1}{2} (\partial \phi_i)^2 - \sum_{i=1}^2 \tfrac{1}{2} e^{2 \phi_i} (\partial \chi_i)^2  - \tfrac{1}{12} e^{-\phi_1 -\phi_2} H_3^2 \nn &&
\phantom{xxxxxxxxxxxx} - \tfrac{1}{12} e^{\phi_1 - \phi_2} F_3^2 - \chi_2 \dd B_2 \wedge \dd C_2,
\eea
where $H_3 = \dd B_2$ and $ F_3 = \dd C_2 - \chi_1 \dd B_2$.  Some sign changes relative to \cite{Duff:1998cr} follow from the difference in conventions. The equations of motion are:
\bea
\label{6deom1} && \dd \left( e^{\phi_1 - \phi_2} *_6 F_3 \right)  - \dd \chi_2 \wedge \dd B_2 = 0, \\
\label{6deom2} && \dd \left( e^{- \phi_1 - \phi_2} *_6 H_3 \right) - e^{\phi_1 - \phi_2} \dd \chi_1 \wedge *_6 F_3 + \dd \chi_2 \wedge F_3 = 0, \\
\label{6deom3} && \dd ( e^{2 \phi_1} *_6 d \chi_1) + e^{\phi_1 - \phi_2} \dd B_2 \wedge *_6 F_3 = 0, \\
\label{6deom4} && \dd (e^{2 \phi_2} *_6 d \chi_2) - \dd B_2 \wedge \dd C_2 = 0, \\
\label{6deom5} && \dd *_6 d \phi_1 - e^{2 \phi_1} \dd \chi_1 \wedge *_6 \dd \chi_1 + \tfrac{1}{2} e^{-\phi_1 - \phi_2} H_3 \wedge *_6 H_3 - \tfrac{1}{2} e^{\phi_1 - \phi_2} F_3 \wedge * F_3 = 0, \\
\label{6deom6} && \dd *_6 d \phi_2 - e^{2 \phi_2} \dd \chi_2 \wedge *_6 \dd \chi_2 + \tfrac{1}{2} e^{-\phi_1 - \phi_2} H_3 \wedge *_6 H_3 + \tfrac{1}{2} e^{\phi_1 - \phi_2} F_3 \wedge * F_3 = 0, \\
\label{6deom7} && R_{\mu \nu} = \tfrac{1}{2} \sum_{i=1}^2 \left( \partial_{\mu} \phi_i \partial_{\nu} \phi_i + e^{2 \phi_i} \partial_{\mu} \chi_i \partial_{\nu} \chi_i \right) \nn && \phantom{xxx} +\tfrac{1}{4} e^{- \phi_1 - \phi_2} \left( H_{3 \mu \rho_1 \rho_2} H_{3\nu}^{~~\rho_1 \rho_2} - \tfrac{1}{6} g_{\mu \nu} H_{3 \rho_1 \rho_2 \rho_3} H_{3}^{~\rho_1 \rho_2 \rho_3} \right) \nn &&
\phantom{xxx} + \tfrac{1}{4} e^{\phi_1 - \phi_2} \left( F_{3 \mu \rho_1 \rho_2} F_{3\nu}^{~~\rho_1 \rho_2} - \tfrac{1}{6} g_{\mu \nu} F_{3 \rho_1 \rho_2 \rho_3} F_{3}^{~\rho_1 \rho_2 \rho_3} \right) .
\eea

\subsection*{Reduction to three dimensions}
To reduce the above equations of motion to three dimensions we substitute in our six-dimensional space-time Ansatz
\be
\dd s^2_6 = \dd s^2_3 + \tfrac{1}{4} e^{2 U} \dd s^2 (S^2) + \tfrac{1}{4} e^{2 V} (\dd z + P + A_1),
\ee
 and expressions for the three-form field strengths (\ref{6dflux}). From (\ref{6deom1}) and (\ref{6deom2}) we get
\bea
\label{Feom1} && \dd ( e^{\phi_1 - \phi_2+ 2 U+V} g  - 2 \chi_2 \sin \alpha ) = 0, \\
&& \dd (e^{\phi_1 - \phi_2 + V - 2U} * G_1) + \dd \chi_2 \wedge H_2 = 0, \\
&& \dd (e^{\phi_1 - \phi_2 - V + 2U} * G_2 )- 2 e^{\phi_1 - \phi_2 + V - 2U} * G_1 + \tfrac{1}{2} g e^{\phi_1 - \phi_2 + 2U+V} F_2 - \dd \chi_2 \wedge H_1 = 0,\phantom{xxxx}
\eea
and
\bea
\label{Heom1} && \dd (e^{- \phi_1 - \phi_2 + V + 2U} h ) - e^{\phi_1 - \phi_2 + V + 2 U} g \dd \chi_1 + G_0 \dd \chi_2 = 0,\\
&& \dd(e^{-\phi_1 - \phi_2 + V - 2U} * H_1) - e^{-\phi_1 - \phi_2 + V - 2U}  \dd \chi_1 \wedge * G_1 - \dd \chi_2 \wedge G_2 = 0, \\
&& \dd (e^{-\phi_1 - \phi_2 - V + 2U} * H_2) - 2 e^{-\phi_1 - \phi_2 + V - 2U} * H_1 + \tfrac{1}{2} h e^{- \phi_1 - \phi_2+ V + 2U } F_2 \nn
&& \phantom{xxxxxxxxxxx} - e^{\phi_1 - \phi_2 - V + 2U} \dd \chi_1 \wedge * G_2 + \dd \chi_2 \wedge G_1  = 0.
\eea
We can now solve (\ref{Feom1}) and (\ref{Heom1}) to determine $g$ and $h$
\bea
g &=& 2 e^{- \phi_1 + \phi_2 - V - 2U } ( \cos \alpha + \sin \alpha \chi_2), \\
h &=& 2 e^{\phi_1 + \phi_2 - V - 2U} [ \sin \alpha - \cos \alpha \chi_2 + (\cos \alpha + \sin \alpha \chi_2) \chi_1].
\eea
In the process we have chosen the integration constants for convenience.

From (\ref{6deom3}) and (\ref{6deom4}) we get the following two equations:
\bea
&& \dd ( e^{ 2 \phi_1 + 2 U +  V} * \dd \chi_1 ) + \left[ 2 \sin \alpha G_0 e^{\phi_1 - \phi_2 - 2 U - V} - g h e^{\phi_1 - \phi_2 + 2 U + V} \right] \vol_3 \nn &&  \phantom{xxxxxxxx} + e^{\phi_1 - \phi_2 + V - 2U} H_1 \wedge * G_1  + e^{\phi_1 - \phi_2 -V + 2 U} H_2 \wedge * G_2 = 0, \\
&& \dd (  e^{ 2 \phi_1 + 2 U +  V} * \dd \chi_2 )  + \left[ h G_0 - 2 \sin \alpha g \right] \vol_3 + H_1 \wedge G_2 - G_1 \wedge H_2 = 0.
\eea
The final two scalar equations give
\bea
&& \dd ( e^{2 U + V} * \dd \phi_1) - e^{2 \phi_1 +2 U + V} \dd \chi_1 \wedge * \dd \chi_1 + \tfrac{1}{2} e^{-\phi_2-2 U -V } \left[  4 e^{-\phi_1} \sin^2 \alpha - e^{\phi_1} G_0^2 \right] \vol_3   \nn
&& + \tfrac{1}{2} e^{-\phi_2 - 2 U +V} \left[ e^{-\phi_1} H_1 \wedge * H_1 - e^{\phi_1}  G_1 \wedge * G_1 \right]   + \tfrac{1}{2} e^{-\phi_2 +2 U -V} \left[ e^{-\phi_1} H_2 \wedge * H_2 - e^{\phi_1}  G_2 \wedge * G_2 \right] \nn
&& - \tfrac{1}{2} e^{-\phi_2 + 2 U +V} \left[ e^{-\phi_1} h^2 - e^{\phi_1}  g^2 \right] \vol_3 = 0, \\
&& \dd ( e^{2 U + V} * \dd \phi_2) - e^{2 \phi_2 +2 U + V} \dd \chi_2 \wedge * \dd \chi_2 + \tfrac{1}{2} e^{-\phi_2-2 U -V } \left[  4 e^{-\phi_1} \sin^2 \alpha + e^{\phi_1} G_0^2 \right] \vol_3   \nn
&& + \tfrac{1}{2} e^{-\phi_2 - 2 U +V} \left[ e^{-\phi_1} H_1 \wedge * H_1 + e^{\phi_1}  G_1 \wedge * G_1 \right]   + \tfrac{1}{2} e^{-\phi_2 +2 U -V} \left[ e^{-\phi_1} H_2 \wedge * H_2 + e^{\phi_1}  G_2 \wedge * G_2 \right] \nn
&& - \tfrac{1}{2} e^{-\phi_2 + 2 U +V} \left[ e^{-\phi_1} h^2 + e^{\phi_1}  g^2 \right] \vol_3 = 0.
\eea

We now only have to work out the Einstein equation. Taking into account a change in how we define scalars, namely $W \rightarrow V, V_1 \rightarrow U$, we can use the Ricci tensor appearing in (\ref{RicciAconst}). We simply have to take note of the fact that the $S^2$ is normalised so that $l_1 = 4$, in which case $A=0$.

From the Einstein equation, we get the following equations:
\bea
&& 2 e^{2 V -4 U} \vol_3 - \dd * \dd V - \dd V \wedge * \dd V - 2 \dd V \wedge * \dd U + \tfrac{1}{8} e^{2V} F_2 \wedge * F_2 \nn
&& = \left[ \tfrac{1}{4} e^{- \phi_2 -2 V -4 U} ( 4 e^{-\phi_1} \sin^2 \alpha+ e^{\phi_1} G_0^2 ) +   \tfrac{1}{4} e^{- \phi_2} ( e^{-\phi_1} h^2 + e^{\phi_1} g^2 )\right] \vol_3 \\
&& + \tfrac{1}{4} e^{-\phi_2-2 V} \left[ e^{- \phi_1} H_2 \wedge * H_2 + e^{\phi_1} G_2 \wedge * G_2 \right] - \tfrac{1}{4} e^{- \phi_2 -4 U} \left[ e^{- \phi_1} H_1 \wedge * H_1 + e^{\phi_1} G_1 \wedge * G_1  \right] \nonumber
\eea
\bea
&& (4 e^{-2 U} - 2 e^{2V -4U}) \vol_3 - \dd * \dd U - \dd U \wedge * \dd V - 2 \dd U \wedge * \dd U \nn
&& = \left[ \tfrac{1}{4} e^{- \phi_2 -2 V -4 U} ( 4 e^{-\phi_1} \sin^2 \alpha+ e^{\phi_1} G_0^2 ) +   \tfrac{1}{4} e^{- \phi_2} ( e^{-\phi_1} h^2 + e^{\phi_1} g^2 )\right] \vol_3 \\
&& - \tfrac{1}{4} e^{-\phi_2-2 V} \left[ e^{- \phi_1} H_2 \wedge * H_2 + e^{\phi_1} G_2 \wedge * G_2 \right] +\tfrac{1}{4} e^{- \phi_2 -4 U} \left[ e^{- \phi_1} H_1 \wedge * H_1 + e^{\phi_1} G_1 \wedge * G_1  \right] \nonumber
\eea
\bea
 \tfrac{1}{2} e^{-2 U - 2V} \dd ( e^{3 V + 2 U}  * F_2) &=&  2 \sin \alpha e^{-\phi_1 - \phi_2 -4 U -V} * H_1 + G_0 e^{\phi_1 - \phi_2 -4 U -V} * G_1 \nn
 &-& e^{- \phi_1 - \phi_2 -V} h H_2 - e^{\phi_1 - \phi_2 -V} g G_2.
\eea

\end{document}